\documentclass[11pt,a4paper]{article}
\usepackage[colorlinks=true, linkcolor=black!50!blue, urlcolor=blue, citecolor=blue, anchorcolor=blue]{hyperref}
\usepackage[font=small,labelfont=bf,margin=0mm,labelsep=period,tableposition=top]{caption}
\usepackage[a4paper,top=3cm,bottom=2.5cm,left=2.5cm,right=2.5cm,bindingoffset=0mm]{geometry}

\usepackage{graphicx,placeins}
\usepackage{float}
\usepackage{afterpage}
\usepackage{epsfig,cite}
\usepackage{amssymb}
\usepackage{amsmath}
\usepackage{dsfont}
\usepackage{multirow}
\usepackage{url}
\usepackage{xcolor,colortbl}
\usepackage{float}
\usepackage{afterpage}
\usepackage{url}
\usepackage{hyperref}
\usepackage{booktabs}
\usepackage{mathrsfs}

\newcommand{\cmark}{\cellcolor{blue!20}\textcolor{blue}{\ding{51}}}%
\newcommand{\xmark}{\cellcolor{red!20}\textcolor{red}{\ding{55}}}%
\newcommand{\ymark}{\cellcolor{yellow!30}\textcolor{blue}{(\ding{51})}}%

\usepackage{tikz}
\usepackage{tikz-3dplot}
\usepackage[compat=1.0.0]{tikz-feynman}

\usepackage{enumitem}
\usepackage{hyperref}
\usepackage{cite}

\usetikzlibrary{shapes, arrows}
\usetikzlibrary{decorations.pathreplacing}
\usetikzlibrary{positioning, calc}
\tikzstyle{fitted} = [rectangle, minimum width=5cm, minimum height=1cm, text centered, draw=black, fill=red!30]
\tikzstyle{operations} = [rectangle, rounded corners, minimum width=2cm,text centered, draw=black, fill=red!30]
\tikzstyle{roundtext} = [rectangle, rounded corners, minimum width=2cm, minimum height=0.8cm, text centered, draw=black, fill=red!30]
\tikzstyle{n3py} = [rectangle, rounded corners, minimum width=3cm, minimum height=1cm, text centered, draw=black, fill=green!30]
\tikzstyle{myarrow} = [thick,->,>=stealth]
\tikzstyle{line} =[draw, -latex']
\tikzstyle{decision} = [diamond, draw, fill=red!20, text width=7.5em, text centered,  inner sep=0pt, minimum height=2em, aspect=4]
\tikzstyle{cloud} = [draw, ellipse,fill=green!20, minimum height=2em]
\tikzstyle{inout} = [rectangle, draw, fill=green!20, text width=9.5em, text centered, rounded corners, minimum height=2em, minimum width=10em]
\tikzstyle{block}=[rectangle, draw, fill=blue!20, text width=9.5em, 
                   text centered, rounded corners, minimum height=2em, 
                   minimum width=10em]

\definecolor{darkgreen}{rgb}{0.0, 0.5, 0.13}

\usepackage[normalem]{ulem}
\definecolor{ao}{rgb}{0.0, 0.5, 0.0}

\newcommand{\be}{\begin{equation}}
\newcommand{\ee}{\end{equation}}
\newcommand{\bea}{\begin{eqnarray}}
\newcommand{\eea}{\end{eqnarray}}
\newcommand{\bi}{\begin{itemize}}
\newcommand{\ei}{\end{itemize}}
\newcommand{\ben}{\begin{enumerate}}
\newcommand{\een}{\end{enumerate}}

\newcommand{\lp}{\left(}
\newcommand{\rp}{\right)}

\def\gsim{\mathrel{\rlap{\lower4pt\hbox{\hskip1pt$\sim$}}
    \raise1pt\hbox{$>$}}}         
\def\lsim{\mathrel{\rlap{\lower4pt\hbox{\hskip1pt$\sim$}}
    \raise1pt\hbox{$<$}}}         

\newcommand{\draft}[1]{}

\def\beq{\begin{equation}}
\def\eeq{\end{equation}}

\def\lapprox{\lower .7ex\hbox{$\;\stackrel{\textstyle <}{\sim}\;$}}
\def\gapprox{\lower .7ex\hbox{$\;\stackrel{\textstyle >}{\sim}\;$}}

\numberwithin{equation}{section}
\numberwithin{figure}{section}
\numberwithin{table}{section}

\usepackage{tabularx}
\newcolumntype{C}[1]{>{\centering\arraybackslash}p{#1}}

\usepackage{amsmath}
\usepackage{amsfonts}
\usepackage{amssymb}
\usepackage{dsfont}
\usepackage{pifont}
\usepackage{booktabs}
\usepackage{graphicx}
\usepackage{epstopdf}
\usepackage{epsfig}
\usepackage{framed}
\usepackage{makeidx}
\usepackage{siunitx}
\usepackage[capitalise]{cleveref}
\usepackage{hyperref}
\usepackage{placeins}
\usepackage[font=small,labelfont=bf]{caption}

\RequirePackage{csquotes}
\usepackage{bbm}

\begin{document}
\newgeometry{top=1.5cm,bottom=1.5cm,left=2cm,right=2cm,bindingoffset=0mm}

\vspace{-2.0cm}
\begin{flushright}
Nikhef-2023-010\\
TIF-UNIMI-2023-27\\
Edinburgh 2023/28
\end{flushright}
\vspace{0.3cm}

\begin{center}
  {\Large \bf Heavy Quarks in Polarised Deep-Inelastic Scattering \\[0.3cm]
    at the Electron-Ion Collider }
  \vspace{1.1cm}

Felix Hekhorn$^{1,2,3}$,
Giacomo Magni$^{4,5}$,
Emanuele R. Nocera$^6$,
Tanjona R. Rabemananjara$^{4,5}$,\\[0.2cm]
Juan Rojo$^{4,5}$, 
Adrianne Schaus$^{4,5}$, and Roy Stegeman$^7$

\vspace{0.7cm}
 
 {\it \small

  ~$^1$Tif Lab, Dipartimento di Fisica, Universit\`a di Milano and\\
   INFN, Sezione di Milano, Via Celoria 16, I-20133 Milano, Italy\\[0.1cm]
   ~$^2$University of Jyvaskyla, Department of Physics, P.O. Box 35, FI-40014 University of Jyvaskyla, Finland\\[0.1cm]
   ~$^3$Helsinki Institute of Physics, P.O. Box 64, FI-00014 University of Helsinki, Finland\\[0.1cm]
   ~$^4$Department of Physics and Astronomy, Vrije Universiteit, NL-1081 HV Amsterdam\\[0.1cm]
   ~$^5$Nikhef Theory Group, Science Park 105, 1098 XG Amsterdam, The Netherlands\\[0.1cm]
   ~$^6$ Dipartimento di Fisica, Universit\`a degli Studi di Torino and\\
   INFN, Sezione di Torino, Via Pietro Giuria 1, I-10125 Torino, Italy\\[0.1cm]
   ~$^7$The Higgs Centre for Theoretical Physics, University of Edinburgh,\\
   JCMB, KB, Mayfield Rd, Edinburgh EH9 3JZ, Scotland\\[0.1cm]
 }

 \vspace{0.7cm}

 {\bf \large Abstract}

\end{center}

We extend the FONLL general-mass variable-flavour-number scheme to the case of
longitudinally polarised DIS structure functions, accounting for perturbative
corrections up to $\mathcal{O}\lp \alpha_s^2\rp$. We quantify the impact of
charm quark mass and higher-order perturbative corrections on projected
measurements of inclusive and charm-tagged longitudinal asymmetries at the
Electron-Ion Collider (EIC) and at the Electron-ion collider in China (EicC).
We demonstrate how the inclusion of these corrections is essential to compute
predictions with an accuracy that matches the projected precision of the
measurements. The computation is made publicly available through the
open-source {\sc\small EKO} and {\sc\small YADISM} programs.

\clearpage
 
\tableofcontents

\section{Introduction}
\label{sec:intro}

The production of charm quarks in unpolarised deep-inelastic scattering (DIS)
contributes significantly to the inclusive structure functions measured by 
HERA~\cite{H1:2015ubc,H1:2018flt}. In particular, it can amount to up
to 25\% at small values of the proton momentum fraction $x$ and at small to
moderate values of the momentum transfer $Q^2$. The accurate determination of
parton distribution functions (PDFs)~\cite{Gao:2017yyd,Kovarik:2019xvh,
  Ethier:2020way} from experimental data therefore requires to include charm
mass effects in the computation of DIS cross sections.
Indeed, all modern PDF determinations~\cite{H1:2015ubc,Alekhin:2017kpj,
  Hou:2019efy,Bailey:2020ooq,NNPDF:2021njg} account for these effects either
with the fixed-flavour-number (FFN) scheme~\cite{Alekhin:2012vu} or with a
general-mass variable-flavour-number (GM-VFN)
scheme~\cite{Thorne:1997uu,Forte:2010ta,Gao:2013wwa}. The latter combines
power-suppressed mass corrections proportional to $m_c^2/Q^2$ with resummation
of collinear logarithms of the form $\ln (Q^2/m_c^2)$, where $m_c$ is the charm
quark mass. GM-VFN schemes provide an accurate description of charm
structure functions for all values of $Q^2$.

The production of charm quarks in polarised DIS is in principle subject to
similar  considerations. Until now, however, a zero-mass variable flavour
number (ZM-VFN)  scheme, whereby charm production is modelled in terms of a
massless charm PDF, has been used in all modern polarised PDF 
determinations~\cite{Nocera:2014gqa,deFlorian:2014yva,DeFlorian:2019xxt,
  Ethier:2017zbq}.
The reason being that charm mass effects are small in the kinematic region
covered by the available polarised DIS datasets, which furthermore are less
precise than their unpolarised counterparts.

This state of affairs will change with the upcoming Electron-Ion 
Collider~\cite{AbdulKhalek:2021gbh}, which is expected to start taking data in
the 2030s. The EIC will be sensitive to polarised DIS structure functions and
asymmetries down to $x\sim 10^{-4}$ for both inclusive and charm-tagged
measurements with unprecedented precision.
Similar considerations  apply to the proposed  Electron-ion collider in China
(EicC)~\cite{Anderle:2021wcy}. The theoretical interpretation of these upcoming
high-precision measurements demands, in analogy with the unpolarised case,
to properly account for charm mass effects and higher-order perturbative
corrections. 

The goal of this paper is to present a unified computational framework in which 
longitudinally polarised structure functions, cross sections, and asymmetries
can be determined using a state-of-the-art treatment of higher-order QCD and
charm-quark corrections.
In particular, this is achieved by extending the FONLL GM-VFN scheme, developed
for unpolarised DIS in~\cite{Forte:2010ta,Ball:2015tna}, to the 
polarised case. The FONLL scheme matches the massive fixed-flavour computation, 
accurate when $Q^2\sim m_c^2$, with the massless computation, accurate when 
$Q^2\gg m_c^2$. Our computational framework is made available through the 
open-source {\sc\small EKO}~\cite{Candido:2022tld} and 
{\sc\small YADISM}~\cite{Candido:2024rkr} software. 

We deploy this framework to evaluate predictions for inclusive and charm-tagged
longitudinally polarised asymmetries in the kinematic region covered by the EIC
and EicC. We specifically quantify the impact of including higher-order and
charm-quark corrections in the computation, and we demonstrate their
comparative relevance to properly match the expected precision of these
measurements. The theoretical accuracy of our framework
therefore represents an important input to analyse future EIC and EicC data, in
particular to determine polarised PDFs.

The outline of this paper is as follows. In Sect.~\ref{sec:nnlo-pol-sfs}
we summarise the  theoretical framework underpinning the calculation of
massless and massive polarised structure functions up to
$\mathcal{O}\lp \alpha_s^2\rp$ accuracy, and discuss their combination into the
FONLL scheme.  In  Sect.~\ref{sec:fonll-predictions} we assess the
phenomenological relevance of heavy quark mass effects and higher-order QCD
corrections on predictions for inclusive and charm-tagged longitudinally
polarised asymmetries at the EIC and EicC. In particular, we compare these
corrections to the projected experimental uncertainties for these observables.
A summary is provided in Sect.~\ref{sec:summary}.
The paper is supplemented with two appendices. Appendix~\ref{app:dglap}
presents a benchmark of the implementation of polarised DGLAP evolution
in {\sc\small EKO} against {\sc\small PEGASUS}~\cite{Vogt:2004ns}.
Appendix~\ref{sec:tmc-impacts} revisits the role 
of target mass corrections in polarised structure functions,
and compares their impact with that
associated with heavy quark and higher-order corrections.

\section{Polarised structure functions in a general-mass scheme}
\label{sec:nnlo-pol-sfs}

In this section we discuss how the FONLL scheme can be extended to the
polarised case. We first review the definition of polarised structure
functions. We then discuss the details of the FONLL scheme in the polarised
case and its implementation in {\sc\small EKO} and {\sc\small YADISM}.
We finally present numerical results in different regions of 
$x$ and $Q^2$ to validate the implementation of the scheme, and
highlight the role played by charm mass effects, by higher-order corrections,
and by the choice of input polarised PDFs.

\subsection{Polarised structure functions revisited}
\label{subsec:pol_obs}

Let us consider lepton-proton polarised DIS where both the lepton and the
proton beams are longitudinally polarised. The differential cross section can
be expressed in terms of the polarised structure functions $g_1$, $g_L$,
and $g_4$ as
\begin{align}
  \begin{split}
    \frac{d^2 \Delta \sigma^j(x,Q^2)}{dx dy} 
    = 
    & \frac{4 \pi \alpha^2}{xyQ^2} \xi^j 
    \left\{ - \left[1 + (1-y)^2\right] g_4^j(x,Q^2) \right.
    \label{eq:diff_xs}\\
    &\hspace{40pt} \left. + y^2 g_L^j(x,Q^2) + (-1)^p 2x \left[1 - (1-y)^2\right] g_1^j(x,Q^2) \right\} \,,	
  \end{split}
\end{align}
with $p= 1$ for leptons and $p = 0$ for anti-leptons. The index $j$
distinguishes charged current (CC) interactions, with $\xi^{\rm CC}=2$,
from neutral current (NC) interactions, with 
$\xi^{\rm NC}=1$. The inelasticity $y$ is given by $y=Q^2/(2x m_N E_{\ell})$ for 
fixed-target scattering and  $y=Q^2/xs$ for collider scattering; $m_N$ is the
proton mass, $E_{\ell}$ is the lepton beam energy, and $s$ is the square of the
centre-of-mass energy. In Eq.~\eqref{eq:diff_xs}, we neglect the polarised
structure functions $g_2$ and $g_3$, which are suppressed by powers of
$W^2/Q^2$, with $W$ being the invariant mass of the hadronic final state.

Provided $Q^2$ is large enough, polarised structure functions can be
factorised as a convolution between perturbative polarised coefficient
functions, $\Delta C^j_{i, k}(x,\alpha_s)$, and non-perturbative,
process-independent polarised PDFs, $\Delta q_k(x,Q^2)$ (for quarks),
and $\Delta g(x,Q^2)$ (for the gluon). These polarised PDFs are defined as the 
difference between the PDFs of partons with the same and with the opposite
helicity as compared to the direction of the proton spin, {\it e.g.} for quarks
\be
\Delta q_k(x,Q^2) = q_k^{\uparrow\uparrow}(x,Q^2) - q_k^{\uparrow\downarrow}(x,Q^2) \, ,
\ee
and likewise for the gluon, where the first arrow indicates the direction of
the proton spin and the second the partonic helicity.
As reviewed in App.~\ref{app:dglap}, these polarised PDFs 
satisfy polarised DGLAP evolution equations in analogy with their unpolarised
counterparts.
At leading twist, this factorised convolution for the polarised structure
functions reads
\begin{align}
  \begin{split}
    g_1^j(x,Q^2) =  \int_{x}^{1} \frac{dz}{z} \Bigg[ \sum_{k={1}}^{n_f} \Delta q_k^+\left(\frac x z,Q^2\right) & \Delta C^j_{1, k}(z,\alpha_s) + \Delta g\left(\frac x z,Q^2\right) \Delta C^j_{1,g}(z,\alpha_s) \Bigg] \,, \\
    g_i^j(x,Q^2) =  \int_{x}^{1} \frac{dz}{z} \Bigg[ \sum_{k={1}}^{n_f} \Delta &q_k^-\left(\frac x z,Q^2\right) \Delta C^j_{i, k}(z,\alpha_s) \Bigg] \, ,\quad i= 4,L \,,
  \end{split}
  \label{eq:sf_convolution}
\end{align}
with $n_f$ the number of active quark flavours and
$\Delta q_k^\pm = \Delta q_k \pm \Delta \bar q_k$
defining the usual sea and valence quark flavour combinations.
Being leading-twist,
Eq.~\eqref{eq:sf_convolution} does not include target mass corrections (TMCs),
which are reviewed in App.~\ref{sec:tmc-impacts}.

The dominant contribution to the double differential cross section
Eq.~\eqref{eq:diff_xs} is provided by the parity-conserving $g_1$ structure
function. Therefore, we will henceforth focus only on this specific structure
function. Furthermore, we restrict ourselves to the electromagnetic case,
in which a virtual photon is exchanged in the hard scattering.
Nevertheless, our discussion can be generalised to the other polarised
structure functions.

Rearranging the quark PDFs in linear combinations which are convenient
for DGLAP evolution, see App.~\ref{app:dglap},
the structure function $g_1$ can be expressed as
\begin{align}
g_1 (x, Q^2) &=
\left ( \frac{1}{n_f} \sum_{k=1}^{n_f} e_{q_k}^2 \right ) \int_{x}^{1} \frac{dz}{z} \Bigg[\Delta\Sigma \left( \frac{x}{z}, Q^2 \right) \Delta C_1^{\rm PS} 
  \left( z, \alpha_s\right) + \Delta g \left( \frac{x}{z} , Q^2 \right) \Delta C_{1,g} 
  \left(z,\alpha_s\right) \Bigg] \nonumber  \\
&\hspace{10pt}  + \sum_{k=1}^{n_f} e_{q_k}^2 
\int_{x}^{1} \frac{dz}{z}  \Delta q_k^+ \left(\frac{x}{z},Q^2\right) 
\Delta C_1^{\rm NS} \left( z, \alpha_s\right) \, ,
\label{eq:leading-twist-g1}
\end{align}
where $e_{q_k}$ is the fractional quark charge.
The polarised structure function $g_1 (x, Q^2)$ 
is therefore decomposed into three contributions proportional to the quark
non-singlet (NS), gluon, and quark pure singlet (PS) coefficient functions.
The latter is defined as the 
difference between the singlet (S) and NS coefficient functions,
$\Delta C_1^{\rm PS} = \Delta C_1^{\rm S} - \Delta C_1^{\rm NS}$.

Equation~\eqref{eq:leading-twist-g1} assumes that all active quarks at the
scale $Q^2$ can be treated as massless. However, quark mass effects cannot be
neglected when the value of $Q^2$ is close to the value of a heavy quark mass
$m_h$. Such effects can be included by modifying the
expressions for the coefficient functions, so that $g_1$ reads as
\begin{align}
\begin{split}
  g_1 (x, Q^2, m_h^2) &= \left ( \frac{1}{n_f} \sum_{k=1}^{n_f} e_{q_k}^2 \right ) \int_{x}^{1} \frac{dz}{z}  \Bigg[\Delta\Sigma \left( \frac{x}{z}, Q^2 \right) \Delta C_1^{\rm PS} 
    \left( z, \alpha_s, \frac{m_h^2}{Q^2}\right) \\
    &\hspace{120pt} + \Delta g \left( \frac{x}{z} , Q^2 \right) \Delta C_{1,g} 
    \left(z,\alpha_s,\frac{m_h^2}{Q^2}\right) \Bigg]\\
  &\hspace{10pt} +  \sum_{k=1}^{n_f} e_{q_k}^2 
  \int_{x}^{1} \frac{dz}{z}  \Delta q_{k}^+ \left(\frac{x}{z},Q^2\right) 
  \Delta C_1^{\rm NS} \left( z, \alpha_s,\frac{m_h^2}{Q^2}\right) \,.
\end{split}
\label{eq:leading-twist-g1-massive}
\end{align}
The polarised structure function is then recast into light, heavy,
and light-heavy contributions
\be
g_1(x,Q^2,m_h^2) = g_{1}^{(\ell)}(x,Q^2) + g_{1}^{(h)}(x,Q^2,m_h^2/Q^2) + g_{1}^{(\ell h)}(x,Q^2,m_h^2/Q^2) \, ,
\label{eq:g1total_def}
\ee
where $g_{1}^{(\ell)}$ indicates the contributions from
diagrams where only light quark lines are present, $g_{1}^{(h)}$ those from
diagrams where the heavy quark couples to the virtual gauge boson,
and $g_{1}^{(\ell h)}$ those which contain heavy quark lines but where a light
quark couples to the virtual boson.

The separation between light and heavy structure functions in
Eq.~\eqref{eq:g1total_def} is hence affected by an ambiguity concerning in
which category one should assign the $g_{1}^{(\ell h)}$ contribution, involving
heavy quarks in the final state but where only light quarks couple to the
virtual boson. This ambiguity is irrelevant for the inclusive
structure function, but it affects the heavy quark structure functions.

The case of charm production is of particular phenomenological interest.
The experimental definition of the charm structure function $g_1^c$ is based on
tagging charm quarks (or charmed hadrons) in the final state, hence it would
include $g_{1}^{(\ell h)}$.
However, the theoretical infrared-safe definition of $g_1^c$
coincides with $g_{1}^{(h)}$, and contains only diagrams where the charm quark
couples with the virtual boson. Here we adopt the  same convention as 
in~\cite{Forte:2010ta}, and define the charm structure function exclusively in
terms of $g_{1}^{(h)}$, while the $g_{1}^{(\ell h)}$ contribution enters only the
total structure function. The latter term is  non-zero only starting at 
$\mathcal{O}\lp \alpha_s^2\rp$ and is small in the region relevant for both
current and future measurements.
In the $n_f=3$ massive scheme, the charm
structure function at the first non-trivial order
is expressed in terms of the gluon polarised PDF,
\begin{equation}
g_1^c (x, Q^2,m_c^2)=   e_c^2\int_{x}^{1} 
\frac{dz}{z} \Delta g \left( \frac{x}{z} , Q^2 \right) \Delta C^{c}_{1,g} 
\left(z,\alpha_s,\frac{m_c^2}{Q^2}\right) \, ,
\label{eq_g1_charm_nlo}
\end{equation}
with the first non-zero term of the gluon coefficient function $\Delta C_{1,g} $
being $\mathcal{O}\lp \alpha_s\rp$.

The massless coefficient functions entering the polarised structure function
$g_1$, Eq.~\eqref{eq:leading-twist-g1}, have been computed at NNLO
in~\cite{Zijlstra:1993sh} and recently at N$^3$LO
in~\cite{Blumlein:2022gpp}.\footnote{For the other polarised 
structure functions, massless coefficient functions were computed at NLO 
in~\cite{deFlorian:1994wp,Anselmino:1996cd} and recently at NNLO 
in~\cite{Borsa:2022irn}.}
The massive coefficient functions entering
Eq.~\eqref{eq:leading-twist-g1-massive} 
are available up to $\mathcal{O}(\alpha_s^2)$~\cite{Hekhorn:2018ywm} 
together with their corresponding asymptotic limit 
$Q^2 \gg m_h^2$~\cite{Behring:2015zaa,Ablinger:2019etw,Behring:2021asx,Blumlein:2021xlc,Bierenbaum:2022biv,Ablinger:2023ahe}.
In Table~\ref{tab:coeff-funcs} we summarise which polarised neutral-current DIS
coefficient functions are available in the literature and which we have
implemented in  {\sc \small YADISM}. For each  perturbative order (NLO, NNLO,
and N$^3$LO)
we indicate the light-to-light (``light''), light-to-heavy (``heavy''),
heavy-to-heavy (``intrinsic''), and ``asymptotic'' ($Q^2 \gg m_h^2$)
contributions.
As we will see next, all the perturbative ingredients required to implement
FONLL at $\mathcal{O}(\alpha_s^2)$ are available and implemented.
Whereas, in principle, Eq.~\eqref{eq_g1_charm_nlo}
could be extended to account for a polarised intrinsic charm component,
as done for the unpolarised case~\cite{Ball:2022qks}, we neglect it here and
set it to zero.
The implementation of the massless polarised
coefficient functions and structure functions in {\sc\small YADISM}
has been benchmarked against {\sc\small APFEL}\cite{Bertone:2013vaa} 
and {\sc\small APFEL++}\cite{Bertone:2017gds}
up to $\mathcal{O}(\alpha_s^2)$, finding satisfactory
agreement~\cite{Candido:2024rkr}.

\begin{table}
\centering
\renewcommand{\arraystretch}{1.50}
\begin{tabular}{lcccc}
\toprule
& light & heavy & intrinsic & asymptotic \\
\midrule
NLO            &  \cmark & \cmark & \xmark & \cmark \\
NNLO         &  \cmark & \cmark & \xmark & \cmark \\
N$^3$LO &  \ymark & \xmark & \xmark & \ymark \\
\bottomrule
\end{tabular}
\vspace{0.3cm}
\caption{Overview of polarised neutral-current DIS coefficient functions
  available in the literature and implemented in {\sc \small YADISM} (blue),
  available in the literature (only for $g_1$), but not implemented in 
  {\sc \small YADISM} (yellow), and not available in the literature (red).
  For each perturbative order (NLO, NNLO, and N$^3$LO) we indicate
  the light-to-light (``light''), light-to-heavy (``heavy''), 
  heavy-to-heavy (``intrinsic''), and ``asymptotic'' 
  ($Q^2 \gg m_h^2$ limit) coefficients functions
  which have been implemented and benchmarked.}
\label{tab:coeff-funcs}
\end{table}

\subsection{The FONLL scheme for polarised structure functions}
\label{sec:fonll-theory}

The FONLL scheme was originally proposed in~\cite{Cacciari:1998it} to account
for heavy quark mass effects in $D$- and $B$-meson production in hadronic
collisions, and was later generalised to unpolarised DIS~\cite{Forte:2010ta},
eventually taking into account
an intrinsic charm contribution~\cite{Ball:2015tna}.
The basic idea underlying FONLL is
best exemplified in the case of charm quark mass effects.
There FONLL combines the massive (three-flavor-number, 3FN) and massless
(four-flavor-number, 4FN) schemes through a suitable matching procedure.
Since both the 3FN and the 4FN schemes are well defined 
factorisation schemes, the FONLL framework has the advantage that it can be
generally applied to any (un)polarised electro- and hadro-production processes
without the need to rely on alternative factorisation schemes.
Whereas henceforth we will focus on charm, the 
discussion can be readily generalised to the case of bottom,
as well as to that of multiple heavy quarks.

In analogy with the unpolarised case, a generic polarised structure function
in the FONLL scheme with four active quarks can be written as:
\begin{equation}
  \label{eq:polarised-fonll}
  g^{\rm FONLL} (x, Q^2) = g^{[4]}(x, Q^2) + g^{[3]}(x, Q^2) - g^{[3,0]}(x, Q^2),
\end{equation}
where the 3FN- and 4FN-scheme structure functions are respectively given by:
\begin{align}
  g^{[3]}(x, Q^2) &= \int_{x}^{1} \frac{dz}{z} \sum_{i=g,q,\bar{q}} \Delta f_i^{[3]} \left( \frac{x}{z},
  Q^2 \right) \Delta C_i^{[3]} \left( z, \alpha_s^{[3]}, \frac{m_c^2}{Q^2} \right) \, ,
  \label{eq:polarised-3fns} \\
  g^{[4]}(x, Q^2) &= \int_{x}^{1} \frac{dz}{z} \sum_{i=g,q,\bar{q},c,\bar{c}} \Delta f_i^{[4]} \left( \frac{x}{z},
  Q^2 \right) \Delta C_i^{[4]} \left( z, \alpha_s^{[4]} \right) \, ,
  \label{eq:polarised-4fns}
\end{align}
with $q$ and $c$ denoting the light quarks and the charm quark, respectively.
The PDFs and strong coupling entering the 3FN structure function in
Eq.~(\ref{eq:polarised-fonll}) can be expressed in terms of their 4FN
counterparts, by means of the matching
relations provided below.
The asymptotic limit ($Q^2\gg m_c^2$) of the massive calculation, $g^{[3,0]}$,
ensures that terms appearing in both the 3FN and 4FN schemes cancel out for
virtualities much higher than that charm quark mass, and it is given by
\begin{align}
  \label{eq:massive0}
  g^{[3,0]}(x, Q^2) = \int_{x}^{1} \frac{dz}{z} \sum_{i=g,q,\bar{q}} \Delta f_i^{[3]} \left( \frac{x}{z},
  Q^2 \right) \Delta C_i^{[3,0]} \left( z, \alpha_s^{[3]}, \log \frac{m_c^2}{Q^2} \right), 
\end{align}
where $ \Delta C_i^{[3,0]}$ is the massless (asymptotic) limit of the polarised
massive coefficient function, in which only the collinear logarithms
$\log (m^2_c/Q^2)$ are retained and mass-suppressed terms are neglected.

As pointed out in~\cite{Forte:2010ta},
there is some flexibility in choosing the perturbative accuracy at which heavy 
quark mass terms are included in the 3FN and 4FN schemes.
In particular, three different variants can be considered: FONLL-A, in which
both 3FN and 4FN expressions are computed at $\mathcal{O}(\alpha_s)$; FONLL-B,
in which the 3FN expression is computed at $\mathcal{O}(\alpha_s^2)$ while the
4FN is computed at $\mathcal{O}(\alpha_s)$; and FONLL-C, in which both 3FN and
4FN expressions are computed at $\mathcal{O}(\alpha_s^2)$. 

It is clear from Eq.~(\ref{eq:polarised-fonll}) that in the asymptotic
limit the FONLL expression reduces
to the 4FN scheme owing to the fact that the difference term
$\left( g^{[3]} - g^{[3,0]} \right)$ vanishes by construction.
On the other hand, in the threshold region $m^2_c\sim Q^2$,
the difference term $g^{[d]} \equiv \left( g^{[4]} - g^{[3,0]} \right)$
vanishes only up to higher-order perturbative corrections, which can be
numerically large.
Different options are available to reduce the impact of a non-vanishing value
of $g^{[d]}$ near the threshold region so that the 3FN calculation is recovered.
One option, known as $\chi$-scaling, consists in replacing
the lower integration limit $x$ in the convolutions entering $g^{[d]}$, namely
Eq.~(\ref{eq:polarised-4fns}) and~(\ref{eq:massive0}),
with a scaling variable $\chi =x(1+4m_c^2/Q^2)$, 
motivated by the physical threshold for charm quark pair production.

In FONLL, one adopts instead a damping prescription, which is based on rewriting
Eq.~(\ref{eq:polarised-fonll}) as
\begin{align}
  g^{\rm FONLL} (x, Q^2) = g^{[3]}(x, Q^2) + D\left( \frac{m^2_c}{Q^2} \right) g^{[d]}(x,Q^2),
  \quad
  D\left( \frac{m^2_c}{Q^2} \right) \equiv \Theta \left(Q^2 - m^2_c \right) \left(1 - \frac{m^2_c}{Q^2} \right)^2\,.
  \label{eq:polarised-fonll-damping}
\end{align}
The definition of the damping factor $D$ in
Eq.~(\ref{eq:polarised-fonll-damping}) ensures that the difference
term $g^{[d]}$, formally of higher order,
is suppressed close to the threshold region $m^2_c \sim Q^2$,
without affecting the required cancellation between $g^{[3]}$ and $g^{[3,0]}$
in the asymptotic limit $Q^2\gg m_c^2$.
In this work, when presenting results for the polarised FONLL structure
functions, we adopt the threshold damping prescription 
Eq.~\eqref{eq:polarised-fonll-damping}.
For unpolarised structure functions,
the numerical impact of this threshold damping prescription is large in FONLL-A,
and otherwise small in FONLL-B and FONLL-C.
In the polarised case, instead, one finds minimal effects of the damping
prescription for all FONLL variants.

The two expressions in Eqns.~(\ref{eq:polarised-3fns})
and~(\ref{eq:polarised-4fns}) are alternative definitions of the polarised
structure functions that depend on the PDFs and 
strong coupling. As mentioned previously, in order to evaluate the FONLL
expression in Eq.~\eqref{eq:polarised-fonll}, the massive 3FN structure
function needs to be 
expressed in terms of $\Delta f_i^{[4]}$ and $\alpha_s^{[4]}$.
The relations between the PDFs and strong coupling in the two schemes are
defined at some fixed matching scale $\mu_c$ and the corresponding results
at a generic scale $Q^2\ne \mu_c^2$ can be 
obtained using the DGLAP evolution equations, see App.~\ref{app:dglap}.
These matching conditions are given by
\begin{align}
  \alpha_s^{[4]} (\mu^2_c) &= \alpha_s^{[3]} (\mu^2_c) + \sum_{n=2}^{\infty} 
  c_n \left( \alpha_s^{[3]} (\mu^2_c)\right)^n, \label{eq:alphas_matching} \\
  \Delta f_i^{[4]} (x, \mu^2_c) &= \int_{x}^{1} \frac{dz}{z} \sum_{j=g,q,\bar{q}} \Delta f_j^{[3]} 
  \left( \frac{x}{z}, \mu_c^2 \right) \Delta K_{ij} \left( z, \alpha_s^{[4]}(\mu_c^2), \frac{\mu_c^2}{m_c^2} \right)
  \label{eq:pdf_matching}\,.
\end{align}
Note that although it is customary to match at the charm mass scale,
$\mu_c=m_c$, this is not required.

The matching coefficients $c_n$ in Eq.~(\ref{eq:alphas_matching}) for the
strong coupling are known up to four loops~\cite{Chetyrkin:1997sg}.
The polarised matching coefficients $\Delta K_{ij}$ in
Eq.~\eqref{eq:pdf_matching} admit a perturbative expansion in $\alpha_s^{[4]}$
whose terms in the series are computed by comparing the computations of the
coefficient functions in the 3FNS and 4FNS.
The components of $\Delta K_{ij}$ for any values of $i$ and $j$ are known up to 
$\mathcal{O}(\alpha_s^2)$~\cite{Bierenbaum:2022biv}. The expression 
of the zeroth order matching coefficients are trivial,
$\Delta K_{ij}^{(0)} = \delta_{ij}$. At $\mathcal{O}(\alpha_s)$, only
$\Delta K_{ij}^{(1)}$ components with $i=g,c,\bar{c}$ and $j=g$ contribute, while
all other components that involve quark lines are nonzero only starting at
$\mathcal{O}(\alpha_s^2)$.

\subsection{Numerical results}
\label{sec:FONLLg1}

The formalism described in Sect.~\ref{sec:fonll-theory} together 
with the theoretical ingredients listed in Sect.~\ref{subsec:pol_obs}
has been implemented in {\sc\small YADISM}, enabling
the calculation of FONLL polarised structure functions at $\mathcal{O}\lp 
\alpha_s\rp$ and $\mathcal{O}\lp \alpha^2_s\rp$. In the following we present 
numerical results for the polarised structure function $g_1$ and $g_1^c$. 
After a review of the features of the current polarised PDF sets,
we check their expected $Q^2$ behaviour, their perturbative
stability, and their dependence on the input polarised PDF set.

\subsubsection{Polarised PDFs}
\label{sec:polpdfs}

\begin{figure}[!t]
\centering
\includegraphics[width=\textwidth]{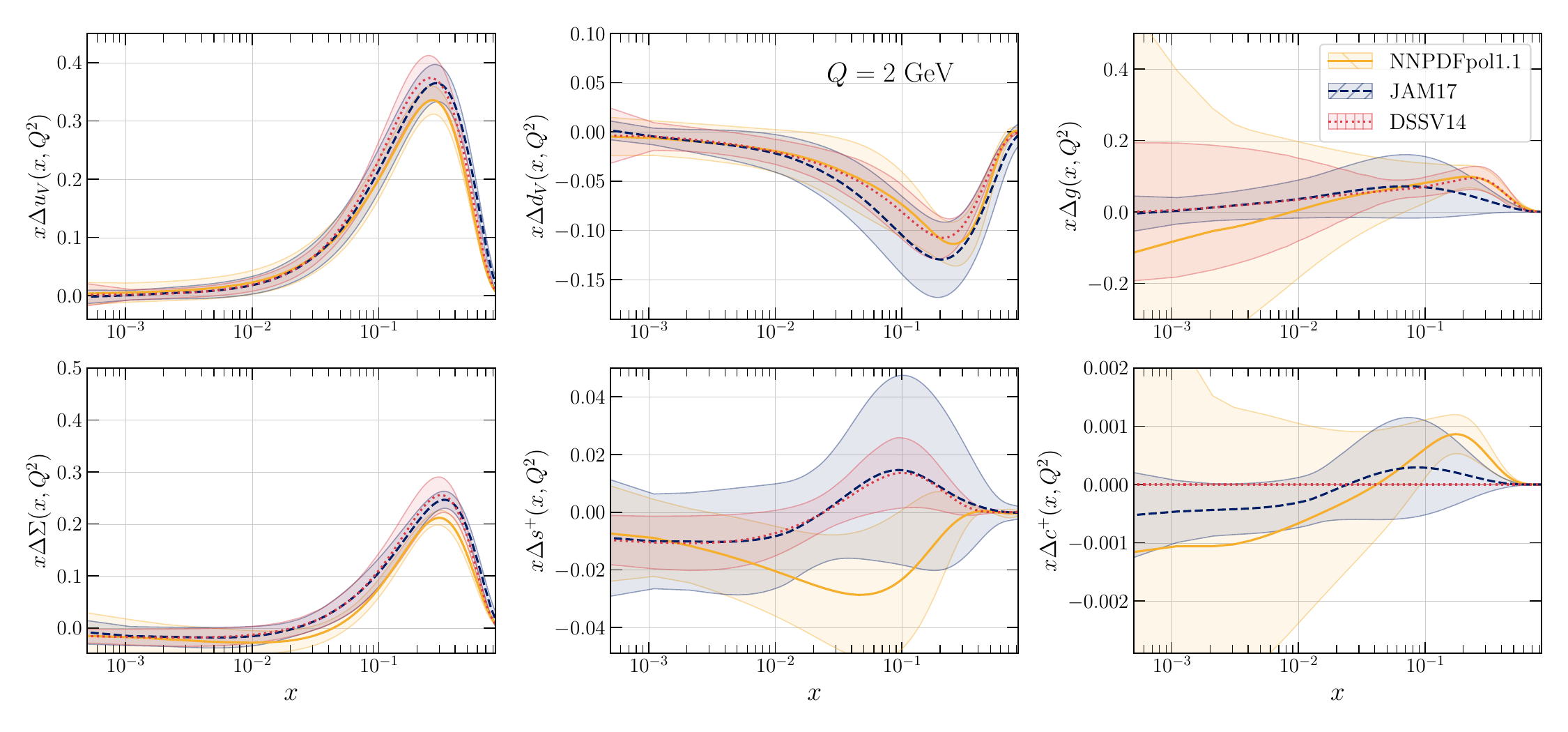}
\caption{Comparison of the polarised proton PDFs from the 
  NNPDFpol1.1~\cite{Nocera:2014gqa}, JAM17~\cite{Ethier:2017zbq}, and
  DSSV14~\cite{DeFlorian:2019xxt} NLO determinations at $Q=2$ GeV. 
  Error bands indicate the corresponding $68\%$ CL PDF 
  uncertainties, evaluated over the Monte Carlo replicas provided by each group.
  We note that the polarised charm PDF has been set to zero in the DSSV14
  Monte Carlo grid.}
\label{fig:PolarisedPDFs-allcomp-abs}
\end{figure}

We present results for FONLL structure functions using alternately two
different determinations of polarised PDFs: NNPDFpol1.1~\cite{Nocera:2014gqa} 
and JAM17~\cite{Ethier:2017zbq}.
These two PDF sets are compared in
Fig.~\ref{fig:PolarisedPDFs-allcomp-abs} at $Q=2$ GeV as a function of $x$,
where the error bands indicate the $68\%$ CL PDF uncertainties.
For completeness, we also include in this comparison
the widely-used DSSV14 polarised PDF set, in particular its Monte Carlo
variant presented in~\cite{DeFlorian:2019xxt}.
We show the up and down valence quarks, gluon, total quark singlet,
strangeness, and charm polarised PDFs.
In DSSV14, the fit is performed in a ZM-VFN scheme but the resulting charm PDF
is set to zero in the released {\sc\small LHAPDF} grids.

Three observations are relevant in light of the subsequent discussion.
First, polarised PDFs are suppressed at small $x$, in contrast with
their unpolarised counterparts in the singlet sector, implying that in general
spin asymmetries (defined as ratios of polarised over unpolarised observables)
are strongly suppressed in this small-$x$ region.
Second, while there is a broad agreement between  the three groups
considered for $\Delta u_V$, $\Delta d_V$, and $\Delta \Sigma$,
there are larger differences for the $\Delta g$, $\Delta s^+$ and $\Delta c^+$.
In particular, the polarised gluon PDF (which drives perturbative charm
production) is poorly known at small $x$ and displays large uncertainties
which then feed into the polarised charm PDF.
Third, the polarised gluon PDF peaks at higher values of $x$ and with a larger
magnitude in NNPDFpol1.1 as compared to JAM17. The same qualitative behaviour
appears in the polarised charm PDF. 

All of these remarks indicate that the bulk of the PDF dependence of polarised 
structure functions and asymmetries, both inclusive and charm-tagged, will be
related to differences at the level of the gluon and charm polarised PDFs.

\subsubsection{$Q^2$ dependence}
\label{sec:Q2dep}

\begin{figure}[!t]
  \centering
  \includegraphics[width=\textwidth]{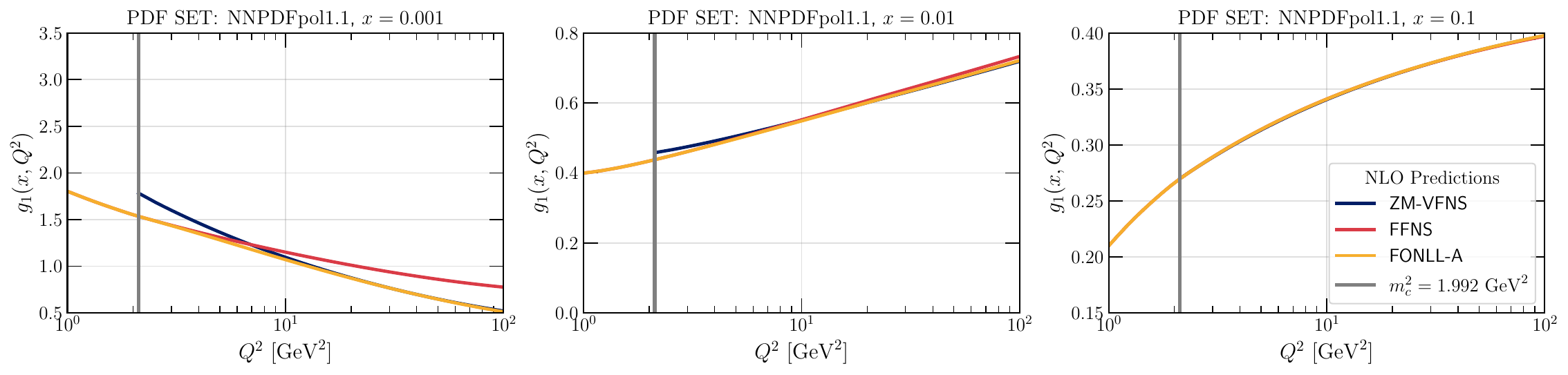}
  \includegraphics[width=\textwidth]{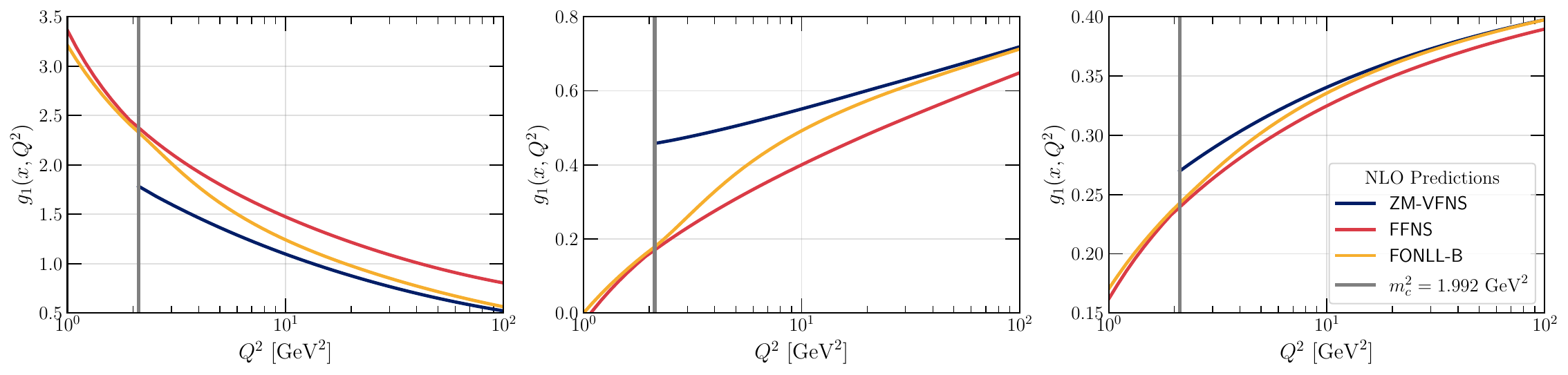}
  \includegraphics[width=\textwidth]{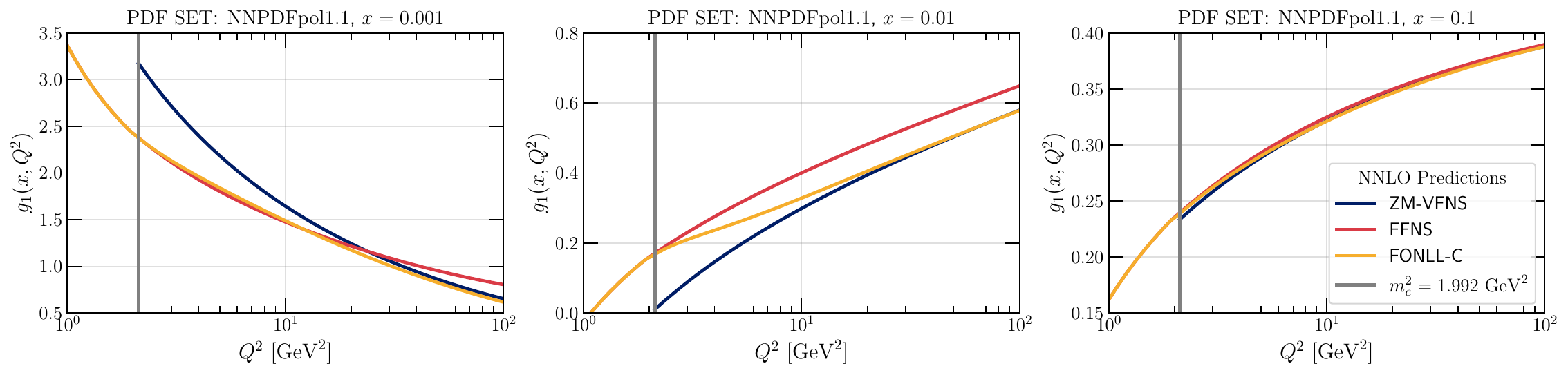}
  \caption{The inclusive polarised structure function $g_1(x,Q^2)$ at three
    fixed values of $x$ ($x=10^{-3},10^{-2}$, and $0.1$) as a function of $Q^2$.
    The central value of the NNPDFpol1.1 NLO polarised PDF set is used as input.
    From top to bottom, we display results corresponding to the FONLL-A, -B, 
    and -C calculations. In each plot, we also display results
    obtained in the ZM-VFN and massive 3FN schemes.
    The vertical grey line indicates the value of $m_c^2$ at which the 3FN and
    4FN schemes are matched.}
\label{fig:g1_nnpdfpol_nnlo_inclusive}
\end{figure}

\begin{figure}[!t]
  \centering
  \includegraphics[width=\textwidth]{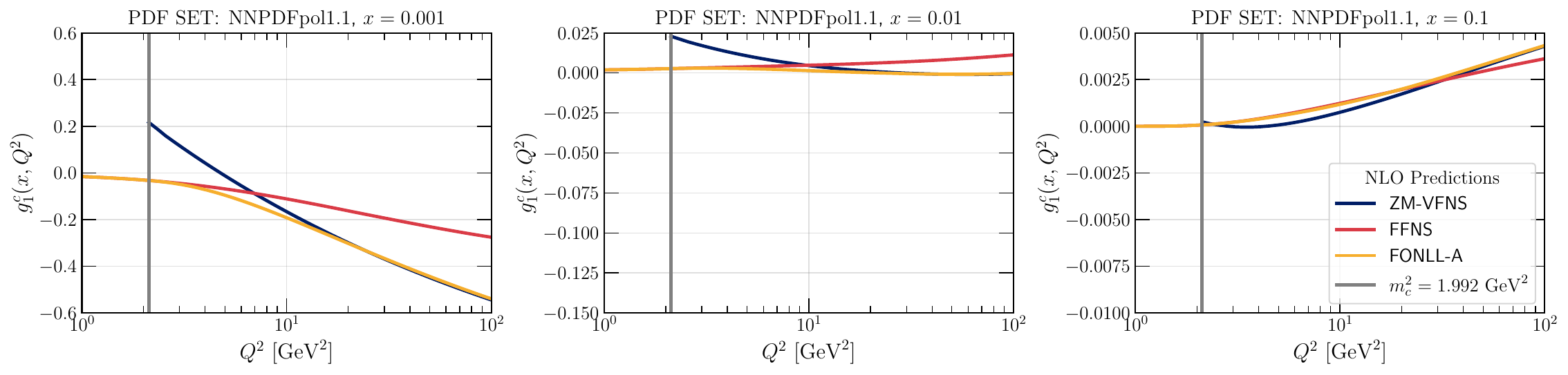}
  \includegraphics[width=\textwidth]{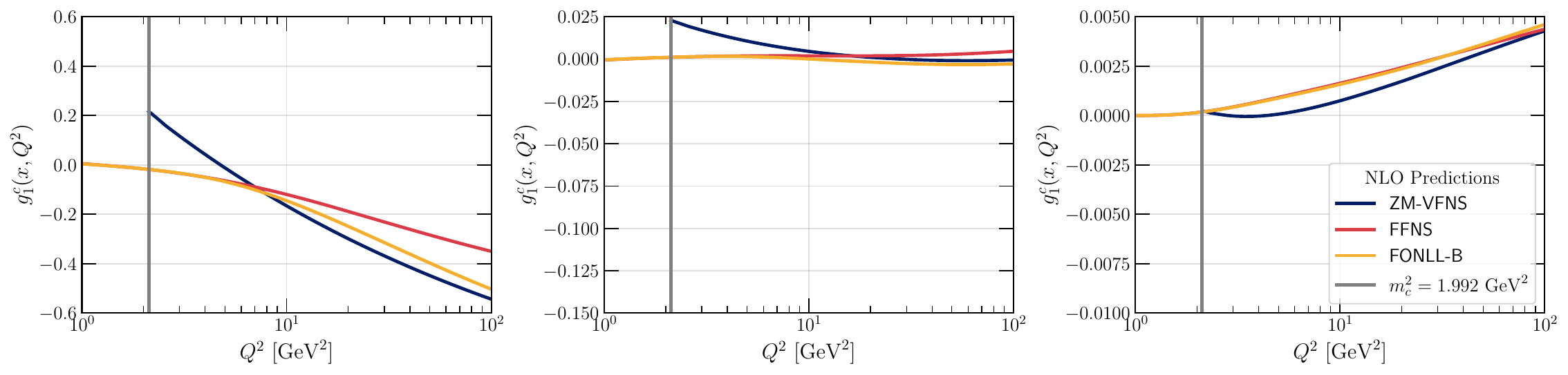}
  \includegraphics[width=\textwidth]{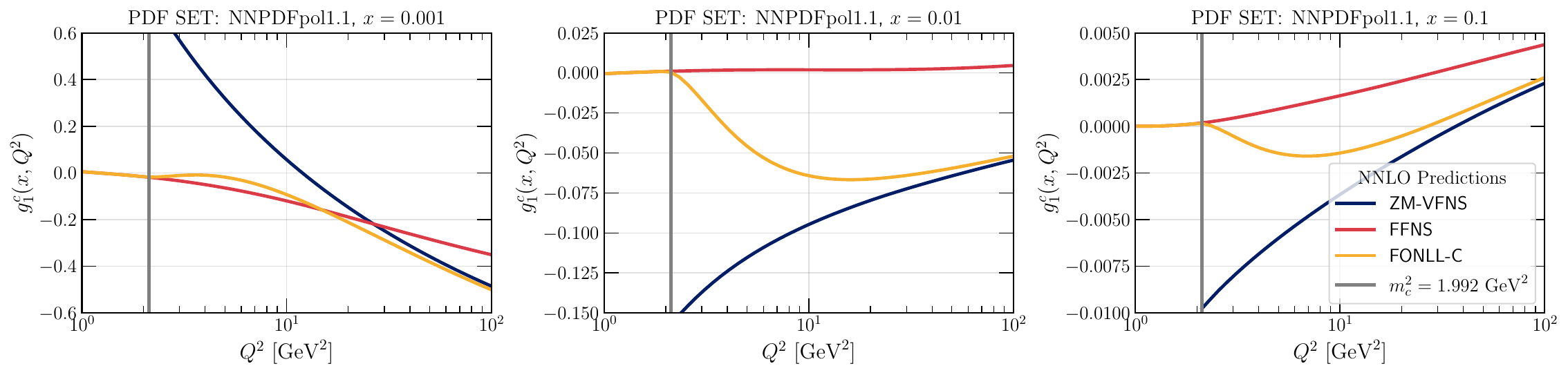}
  \caption{Same as Fig.~\ref{fig:g1_nnpdfpol_nnlo_inclusive} for the charm
    polarised structure function $g_1^c$.}
\label{fig:g1c_nnpdfpol_nnlo_charm}
\end{figure}

Figures.~\ref{fig:g1_nnpdfpol_nnlo_inclusive} 
and~\ref{fig:g1c_nnpdfpol_nnlo_charm} display respectively the inclusive and 
charm polarised structure functions, $g_1(x,Q^2)$ and $g_1^c(x,Q^2)$,
for three fixed values of $x$ ($x=10^{-3},10^{-2},$ and $0.1$) as a function of 
$Q^2$. The central value of the NNPDFpol1.1 NLO polarised PDF set is used as 
input. From top to bottom, we display results corresponding to the FONLL-A, -B, 
and -C.
By construction, the first two are accurate to NLO 
($\mathcal{O}\lp \alpha_s\rp$-accurate), while the last is accurate to NNLO
($\mathcal{O}\lp \alpha_s^2\rp$-accurate). In each plot, we also display results
obtained in the ZM-VFN (only for $Q^2\ge m_c^2$) and massive 3FN schemes. 
The vertical grey line indicates the value of $m_c^2$ at which the 3FN and 4FN 
schemes are matched. 

From these comparisons, one verifies that the FONLL calculation interpolates 
between the massive calculation at low $Q^2$  (close to the charm mass)
and the massless calculation valid for large $Q^2\gg m_c^2$.
For both $g_1$ and
$g_1^c$, charm mass effects can be significant at a scale close to the value
of the charm quark mass. For $g_1$, at $x\sim 10^{-3}$ and $Q^2 = m_c^2$, 
the massless NLO (NNLO) calculation overestimates the matched FONLL 
calculation by up to 15\% (25\%). For $g_1^c$, mass effects cannot be neglected 
until relatively large $Q^2$, given  that only for $Q^2 \gsim 50~\rm{GeV}^2$
the FONLL calculation converges to the massless one.
Interestingly, this holds true also for relatively large $x$ values, such as
$x=0.1$, though in this region $g_1^c$ is relatively small in absolute terms.
From Fig.~\ref{fig:g1c_nnpdfpol_nnlo_charm} one also notes that, depending on 
the value of $x$ and on the perturbative order, the matched FONLL calculation 
deviates from the 3FN scheme calculation already for moderate values of $Q^2$, 
indicating how a purely massive calculation will in general be inadequate to 
describe data unless close to threshold.
The behaviour of this near-threshold 
region exhibits in general a very mild dependence on the choice made for the
damping of subleading terms, see Sect.~\ref{sec:fonll-theory}.

Overall, we conclude that in the kinematic region defined by
$Q^2 \lsim 30$ GeV$^2$ charm quark mass effects cannot be neglected in the 
computation of either the inclusive or charm structure functions.
For larger $Q^2\gsim 30$ GeV$^2$ values, instead, the massless and FONLL
calculations coincide.
This said, it is important to emphasise that a partial
cancellations of charm mass effects may occur if the polarised structure
functions are normalised to their unpolarised counterparts, as happens in the
definition of the experimentally measured spin asymmetries.
We will revisit this issue in Sect.~\ref{sec:fonll-predictions},
where we will compare heavy quark mass effects in inclusive and charm-tagged 
spin asymmetries to the projected precision of EIC and EicC pseudodata.

\subsubsection{Perturbative stability and PDF dependence}
\label{sec:pert_stability}

\begin{figure}[!t]
  \centering
  \includegraphics[width=1\textwidth]{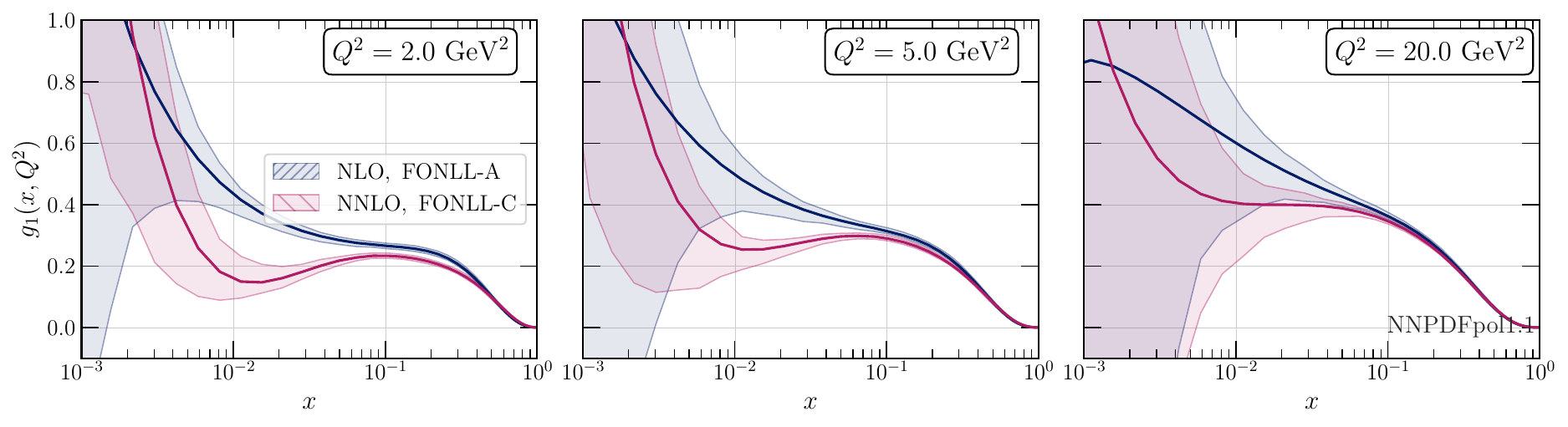}
  \includegraphics[width=1\textwidth]{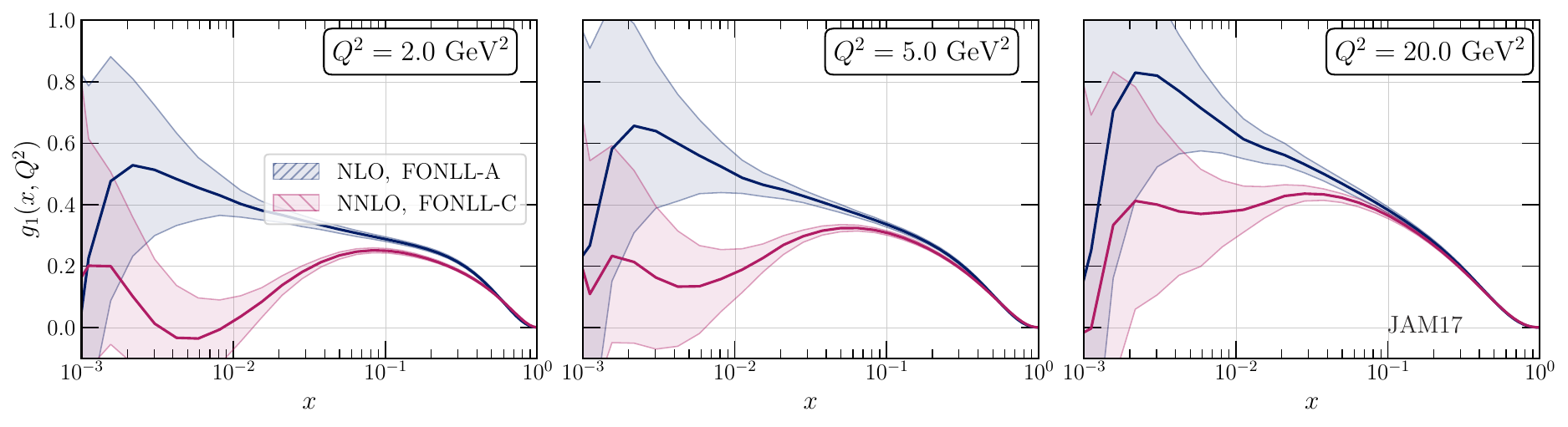}
  \caption{Comparison between the FONLL-A and FONLL-C calculations
    of the inclusive polarised structure function $g_1(x,Q^2)$.
    We display results for NNPDFpol1.1 (top panels) and JAM17 (bottom panels)
    as a function of $x$ for three different values of $Q^2$.
    Error bands correspond to $68\%$ CL PDF uncertainties.}
  \label{fig:g1_total_pto}
\end{figure}

\begin{figure}[!t]
  \centering
  \includegraphics[width=1\textwidth]{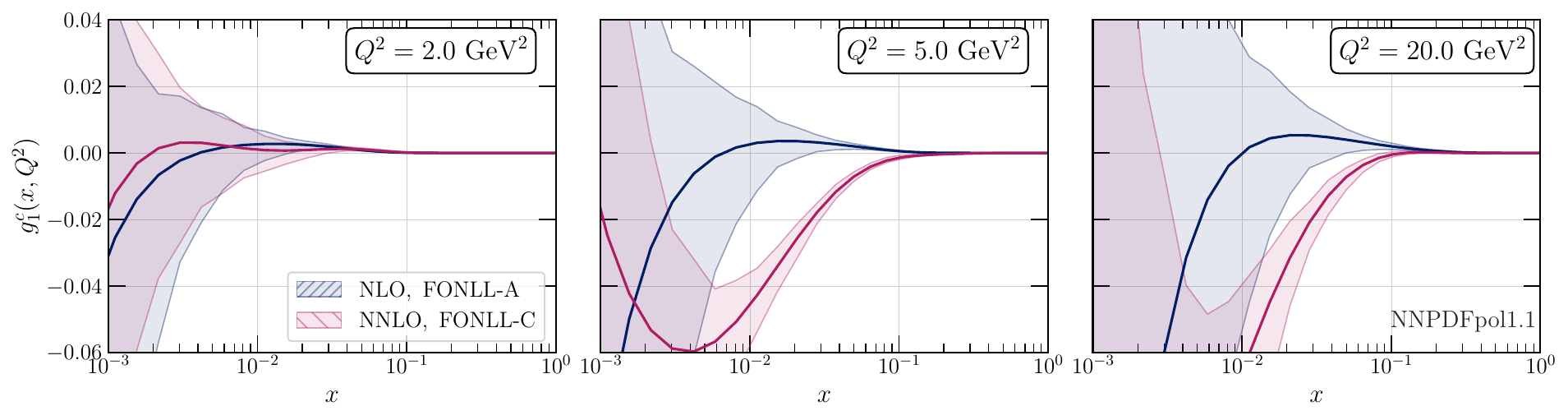}
  \includegraphics[width=1\textwidth]{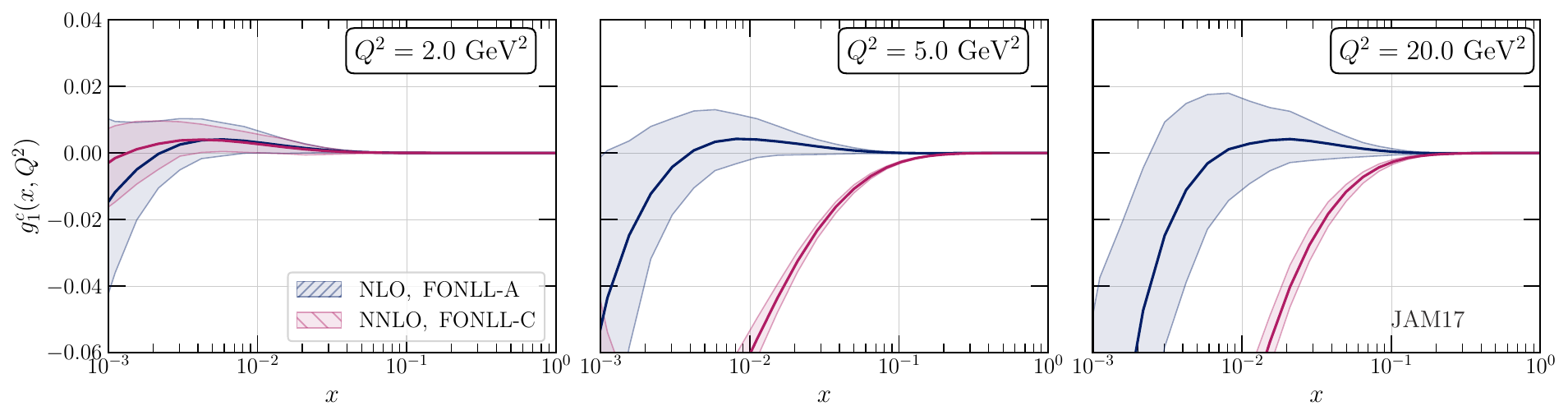}
  \caption{Same as Fig.~\ref{fig:g1_total_pto} for the charm structure function $g_1^c(x,Q^2)$.}
  \label{fig:g1_charm_pto}
\end{figure}

The FONLL structure functions displayed in 
Figs.~\ref{fig:g1c_nnpdfpol_nnlo_charm}
and~\ref{fig:g1_nnpdfpol_nnlo_inclusive} exhibit a clear dependence on
the perturbative accuracy of the calculation.
To showcase these differences in a more direct manner,
Figs.~\ref{fig:g1_total_pto} and~\ref{fig:g1_charm_pto}
display a comparison between the FONLL-A (NLO) and FONLL-C (NNLO) 
calculations for the inclusive $g_1(x,Q^2)$ and charm-tagged $g_1^c(x,Q^2)$ 
polarised structure functions, respectively, for three different values of
$Q^2$ near the matching scale. In both cases, the top and bottom panels show
the predictions using, respectively, the NNPDFpol1.1 and JAM17 polarised PDF
sets as input. Error bands correspond to $68\%$ CL PDF uncertainties.

Figs.~\ref{fig:g1_charm_pto} and~\ref{fig:g1_total_pto} indicate 
that NNLO corrections to both $g_1(x,Q^2)$ and $g_1^c(x,Q^2)$ 
are moderate for $x\gsim 0.1$ and sizeable for $x\lsim 0.05$, 
irrespective of the PDF set used.
Concerning the inclusive structure function $g_1$, the NNLO (FONLL-C)
computation leads to a suppression of the structure function with respect to its
NLO (FONLL-A) counterpart.
Concerning the charm structure function $g_1^c$, the NNLO (FONLL-C) 
computation leads to a negative and large (in absolute value) structure function
in the near-threshold region, whereas the NLO (FONLL-A) computation leaves it 
slightly positive in the same region. Interestingly, NNPDFpol1.1 PDF
uncertainties are sufficiently large to encompass these large differences 
for $x\lsim 0.005$.

The sensitivity of $g_1(x,Q^2)$ and $g_1^c(x,Q^2)$ on the input PDF set
is generally mild: the aforementioned features qualitatively hold when either
the NNPDFpol1.1 or JAM17 PDF sets are used as input.
Small differences are observed, {\it e.g.}\ the shift between FONLL-A and
FONLL-C predictions is larger in JAM17 than in NNPDFpol1.1.
Given that for the sake of the comparison the PDFs have been kept fixed in both
the FONLL-A and -C calculations, one expects that the observed differences may
be reduced once the PDFs are refitted to NNLO accuracy and in the presence 
of the constraints provided by future electron-proton colliders.

\section{Charm mass effects in polarised DIS at electron-ion colliders}
\label{sec:fonll-predictions}

In this section we quantify the impact of charm mass effects on polarised DIS
measurements at future polarised electron-ion colliders, in particular at the
EIC and the EicC. We discuss first the observables and pseudodata sets
considered, then the computation of the corresponding theoretical
predictions, and finally the comparison between the two.
In particular, we assess the significance of charm mass corrections in
comparison to the size of the projected experimental uncertainties
and of the higher-order QCD corrections. 

\subsection{Observables and pseudodata sets}
\label{subsec:pseudodata_sets}

We consider pseudodata sets for the double-spin asymmetry $A_\parallel$
forecast at the EIC, and for the polarised charm asymmetry
$A_1^c$ forecast at the EIC and EicC. The double-spin asymmetry $A_\parallel$
is defined as the ratio of the polarised to unpolarised differential cross
sections,
\begin{equation}
A_\parallel(x,Q^2)
=
\frac{d^2\Delta\sigma(x,Q^2)}{d^2\sigma(x,Q^2)}
=
\frac{d^2\sigma^{\rightarrow\Rightarrow}-d^2\sigma^{\rightarrow\Leftarrow}}{d^2\sigma^{\rightarrow\Rightarrow}+d^2\sigma^{\rightarrow\Leftarrow}}\,,
\end{equation}
where the numerator (denominator) is the difference between (sum of)
differential cross 
sections for which the nucleon is polarised along ($\Rightarrow$) or opposite 
($\Leftarrow$) the polarisation direction of the lepton beam ($\rightarrow$).
Neglecting target mass  corrections, $\mathcal{O}(m_N^2/Q^2)$, which are
expected to be immaterial for EIC and EicC kinematics,\footnote{See
  App.~\ref{sec:tmc-impacts} for the discussion of TMCs.} the asymmetry
$A_\parallel$ becomes proportional to 
the virtual photo-absorption asymmetry $A_1$,
\begin{equation}
A_{||}(x,Q^2) = \mathcal{D}(y) A_1(x,Q^2)\,,
\label{eq:A_parallel}
\end{equation}
where $\mathcal{D}(y)=[y(2-y)]/[y^2+2(1-y)]$ is the photon de-polarisation
factor, and $y$ is the inelasticity. Within the same kinematic approximation,
the photo-absorption asymmetry $A_1$ reads
\begin{equation}
A_1(x,Q^2)=\frac{g_1(x,Q^2)}{F_1(x,Q^2)}\,,
\label{eq:A1}
\end{equation}
where $F_1$ is the unpolarised structure function corresponding to $g_1$.
Likewise, we define the charm photo-absorption asymmetry as
\begin{equation}
A_1^c(x,Q^2)=\frac{g_1^c(x,Q^2)}{F_1^c(x,Q^2)}\,.
\label{eq:A1c}
\end{equation}

For the inclusive double-spin asymmetry $A_{||}$ at the EIC, we use the
projections obtained in~\cite{ATHENA:2022hxb}. These were recently produced in
the context of the performance study of the ATHENA detector, now integrated
into the ePIC detector which will be installed at interaction point IP6 of the
EIC. These projections consider five different beam energy configurations for
electron-proton scattering, each one assuming one year of running:
$5\otimes 41~\mathrm{GeV}$,
$5\otimes 100~\mathrm{GeV}$,
$10\otimes 100~\mathrm{GeV}$,
$10\otimes 275~\mathrm{GeV}$, and
$18\otimes 275~\mathrm{GeV}$,
where the first (second) number indicates the electron (proton) energy.
These five scenarios correspond, respectively, to centre-of-mass energies of
$\sqrt{s}=29$, 45, 63, 105 and 140~GeV, and to integrated
luminosities of $\mathcal{L}=4.4$, 61, 79, 100, and 15.4~fb$^{-1}$.
In all cases, the kinematic coverage considered is $Q^2 \ge 1$ GeV$^2$ and
$0.01 < y < 0.95$. The systematic uncertainties include a point-by-point
uncorrelated systematic uncertainty
(1.5\%),  a normalisation uncertainty (5\%), and a systematic uncertainty of
$10^{-4}$ due to the relative luminosity. Electron and proton beam polarisations
between 70\% and 80\% are assumed.

For the charm photo-absorption asymmetry $A_1^c$, Eq.~(\ref{eq:A1c}), at the
EIC and at the EicC, we use projections from~\cite{Anderle:2021hpa}
and~\cite{Anderle:2023uvi}, respectively.
In the case of the EIC, these projections correspond to three different
beam energy configurations: $5\otimes 41~\mathrm{GeV}$,
$5\otimes 100~\mathrm{GeV}$, and
$18\otimes 275~\mathrm{GeV}$.
The corresponding centre-of-mass energies are $\sqrt{s}=43$, 67,
and 211~GeV. An integrated luminosity of $\mathcal{L}=100$ fb$^{-1}$ is
assumed for all three configurations.
Electron and proton beam polarisations are
of $80\%$ and $70\%$, respectively. In the case of the EicC, projections
correspond to two different beam energy configurations:
$3.5\otimes 20~\mathrm{GeV}$, and
$5\otimes 25~\mathrm{GeV}$. The corresponding centre-of-mass
energy is $\sqrt{s}=15$~GeV and 22~GeV, and the integrated luminosity is
$\mathcal{L}=100$ fb$^{-1}$. For both the EIC and the EicC, the total
experimental uncertainties provided in~\cite{Anderle:2021hpa,Anderle:2023uvi}
include statistical, systematic, and luminosity uncertainties added in
quadrature.

In all cases, we ignore the central values of the aforementioned pseudodata sets
as provided in Refs.~\cite{ATHENA:2022hxb,Anderle:2021hpa,Anderle:2023uvi}.
We retain instead only their projections for the experimental uncertainties as a
function of each bin in $x$ and $Q$.
The projected central values of the pseudodata are then
replaced with our own theoretical predictions, obtained as described next. 

\subsection{Theoretical predictions}
\label{subsec:theoretical_predictions}

We compute theoretical predictions for the inclusive and charm spin asymmetries
corresponding to the pseudodata sets discussed above by using alternately the
ZM-VFN or the FONLL schemes, specifically FONLL-A at NLO and FONLL-C at NNLO,
see Sect.~\ref{sec:nnlo-pol-sfs} for details. We neglect a possible polarised
intrinsic charm component in the proton, TMCs, electroweak corrections, and
corrections due to hadronisation of charm quarks into $D$ mesons. These
corrections are expected to be of similar size in unpolarised and polarised
scattering, therefore they will almost completely cancel in the relevant
asymmetries.
The renormalisation and factorisation scales, $\mu_R$ and $\mu_F$, are set
equal to the DIS virtuality, $\mu_R=\mu_F=Q$.
The same theoretical settings
are adopted consistently in the computation of both the unpolarised and
polarised structure functions entering the asymmetry.

We use the following sets of polarised PDFs: NNPDFpol1.1~\cite{Nocera:2014gqa},
DSSV14~\cite{deFlorian:2014yva,DeFlorian:2019xxt}, and
JAM17~\cite{Ethier:2017zbq} for the computation of inclusive asymmetries;
NNPDFpol1.1 and JAM17 for the computation of charm asymmetries.
In the latter case, we do not use the DSSV14 PDF set because the polarised
charm quark and anti-quark PDFs are identically set to zero in the released
{\sc\small LHAPDF} grid.\footnote{The DSSV14 fit itself adopts a ZM-VFN scheme
  with $n_f^{\rm (max)}=5$ neglecting charm and bottom mass effects.}
By varying the input PDF set one can verify that, whereas predictions
change consistently with Fig.~\ref{fig:PolarisedPDFs-allcomp-abs},
our assessment of the impact of charm-quark mass corrections does not depend on
the specific choice of polarised PDF set. In all cases, we take the NNLO
unpolarised PDF set from the NNPDF4.0 determination~\cite{NNPDF:2021njg}
to evaluate the denominator of the spin asymmetries.

We consider these settings suitable to quantify the role of charm quark mass
effects in EIC spin asymmetries.
They may not necessarily correspond to the
optimal settings that one would adopt to include actual EIC measurements in a
global fit of helicity-dependent PDFs.

\subsection{Comparisons with EIC and EicC projections }
\label{subsec:comparisons}

We now compare the accuracy of theoretical predictions obtained in the ZM-VFN
and FONLL schemes against the expected precision of the pseudodata sets
discussed above.
In particular, we investigate whether differences in the former are larger than
the latter. We discuss in turn the inclusive longitudinal double-spin asymmetry
$A_{||}$ and the charm photo-absorption asymmetry $A_1^c$.

\subsubsection{Inclusive double-spin asymmetry}
\label{subsubsec:aparallel}

\begin{figure}[!t]
  \centering
  \includegraphics[width=\textwidth]{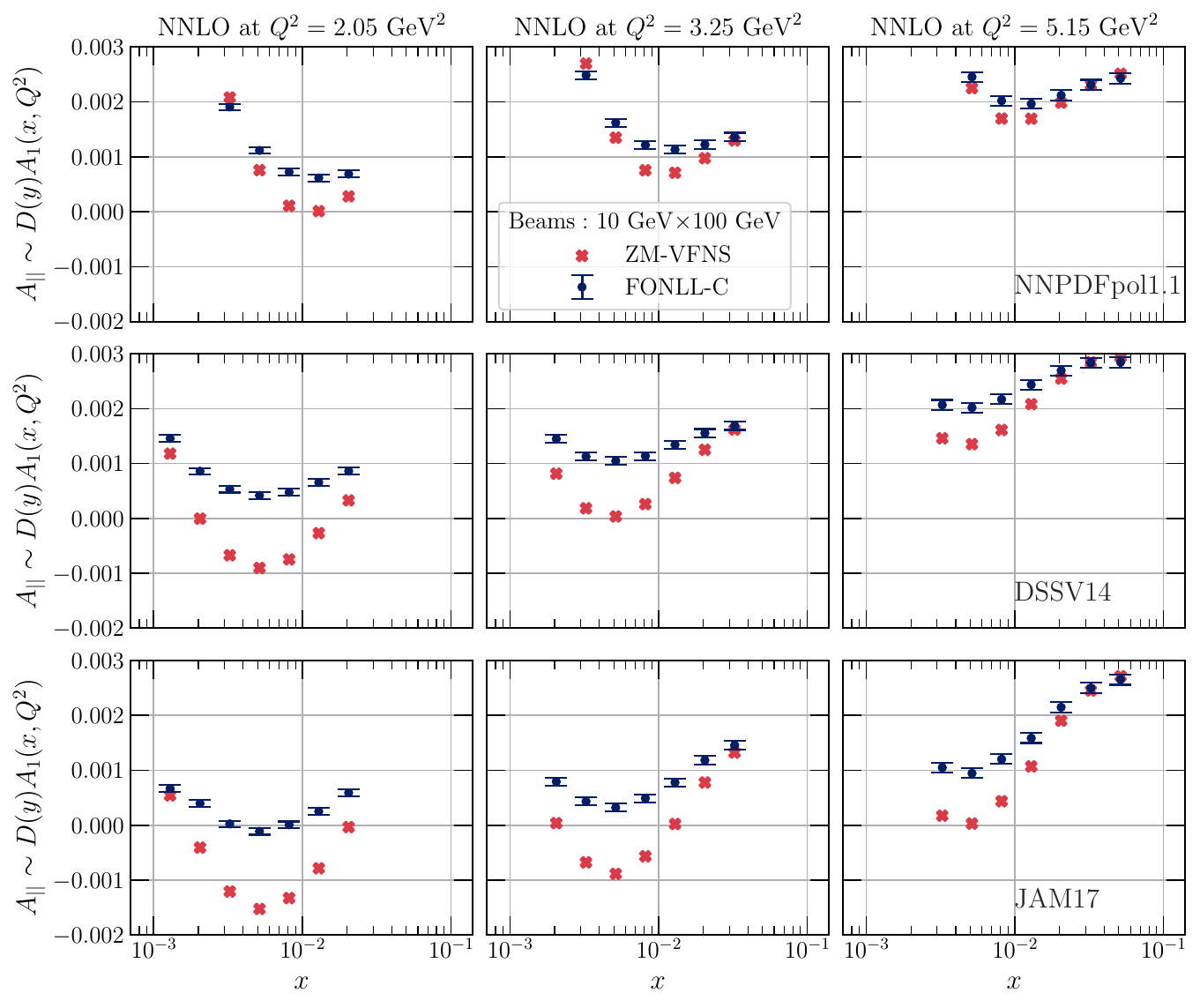}
  \caption{The inclusive longitudinal double spin asymmetry $A_{||}$, defined
    in Eq.~(\ref{eq:A_parallel}), computed at NNLO accuracy with either the
    ZM-VFN or the FONLL-C schemes using the NNPDFpol1.1~\cite{Nocera:2014gqa},
    JAM17~\cite{Ethier:2017zbq}, and DSSV14~\cite{deFlorian:2014yva,
      DeFlorian:2019xxt} polarised PDF sets. Points correspond to a subset of
    the pseudodata discussed in Sect.~\ref{subsec:pseudodata_sets},
    specifically to the low-$Q^2$ bins of the EIC electron-proton beam energy
    configuration 10$\otimes$100~GeV. Error bars, indicated on top of the
    FONLL-C result, correspond to the projected experimental uncertainties.}
  \label{fig:All_athena_projections}
\end{figure}

Fig.~\ref{fig:All_athena_projections} shows the inclusive longitudinal double
spin asymmetry $A_{||}$, Eq.~(\ref{eq:A_parallel}), computed at NNLO accuracy
with either the ZM-VFN or the FONLL-C schemes.
The three aforementioned PDF sets
are used. Error bars, indicated on top of the FONLL-C result, correspond to the
projected experimental uncertainties, see Sect.~\ref{subsec:pseudodata_sets}.
We display only pseudodata points corresponding to the low-$Q^2$ bins of the EIC
electron-proton dataset associated to the beam energy configuration
10$\otimes$100~GeV. For these bins
and this beam energy configuration, quark mass effects are the largest.
We explicitly checked that, at higher values of $Q^2$ or for different
beam energy configurations, the FONLL-C calculation smoothly reduces to the
ZM-VFN.

From Fig.~\ref{fig:All_athena_projections} one observes that predictions
obtained with either the ZM-VFN or the FONLL-C schemes may differ significantly,
especially in the bins with the lowest values of $Q^2$. Predictions obtained
with the former typically undershoot the ones obtained with the latter.
Whereas
the magnitude of the predictions depend on the input polarised PDF set,
especially in the small-$x$  region beyond the coverage of available data,
the impact of charm mass
effects is much larger than the projected experimental uncertainties. As
expected, as $Q^2$ increases, the difference between predictions obtained with
the ZM-VFN or the FONLL-C schemes becomes negligible.
We therefore conclude that
the inclusion of charm mass corrections in the computation of the inclusive
double-spin asymmetry is essential to properly match the forecast EIC
measurements within their precision and robustly interpret them
in terms of the underlying spin decomposition of the proton~\cite{Ball:2013tyh}.

\subsubsection{Charm photo-absorption longitudinal asymmetry}

\begin{figure}[!t]
  \centering
  \includegraphics[width=0.495\textwidth]{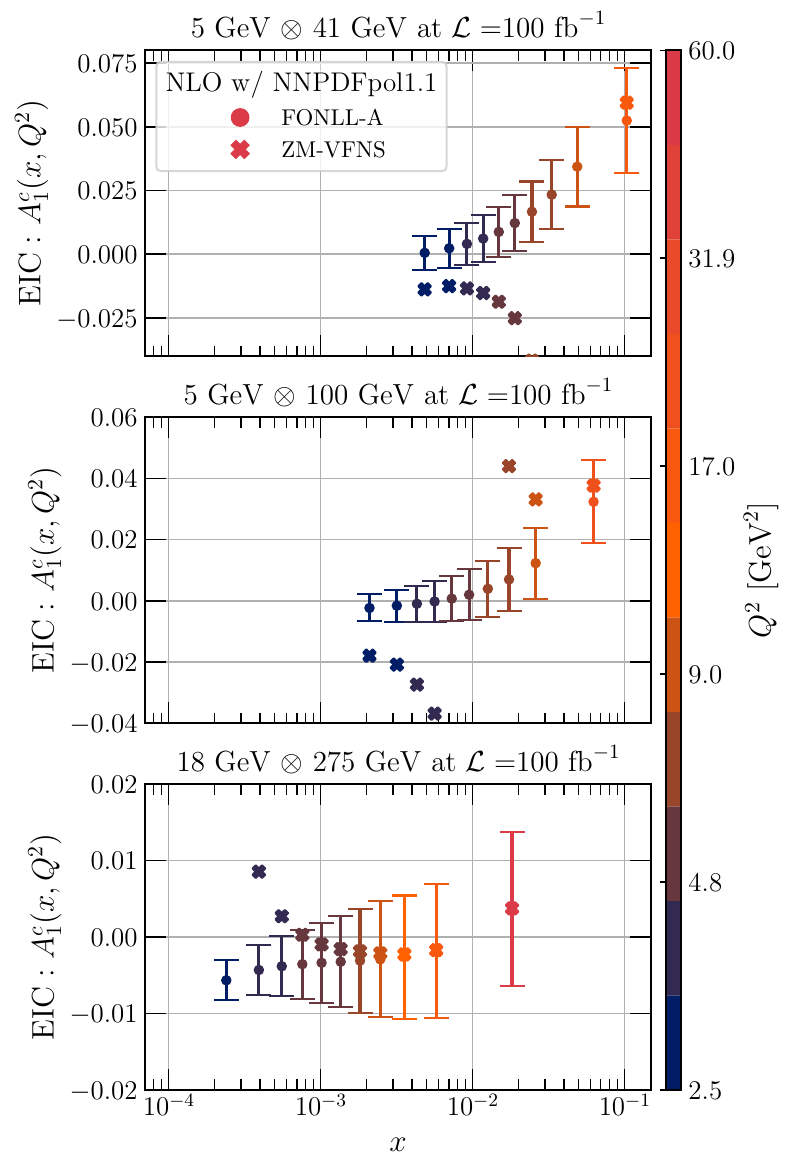}
  \includegraphics[width=0.495\textwidth]{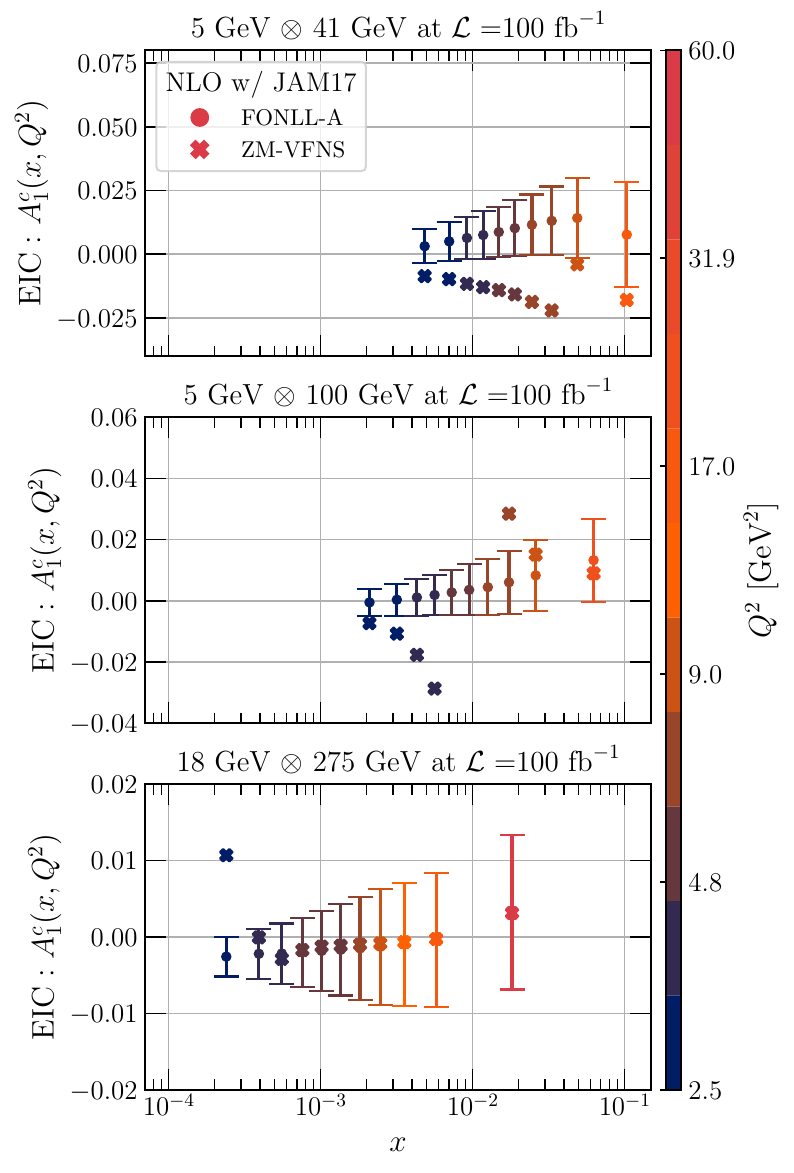}
  \caption{The charm longitudinal asymmetry $A^c_1$, defined in
    Eq.~\eqref{eq:A1c} computed at NLO accuracy with either the ZM-VFN or the
    FONLL-A schemes using the NNPDFpol1.1~\cite{Nocera:2014gqa} (left) and
    JAM17~\cite{Ethier:2017zbq} (right) polarised PDF sets. Points correspond
    to the pseudodata discussed in Sect.~\ref{subsec:pseudodata_sets}, at
    different values of $x$ and $Q^2$. Error bars, indicated on top of the
    FONLL-C result, correspond to the projected experimental uncertainties.}
  \label{fig:a1c_nlo_EIC}
\end{figure}

\begin{figure}[!t]
  \centering
  \includegraphics[width=0.495\textwidth]{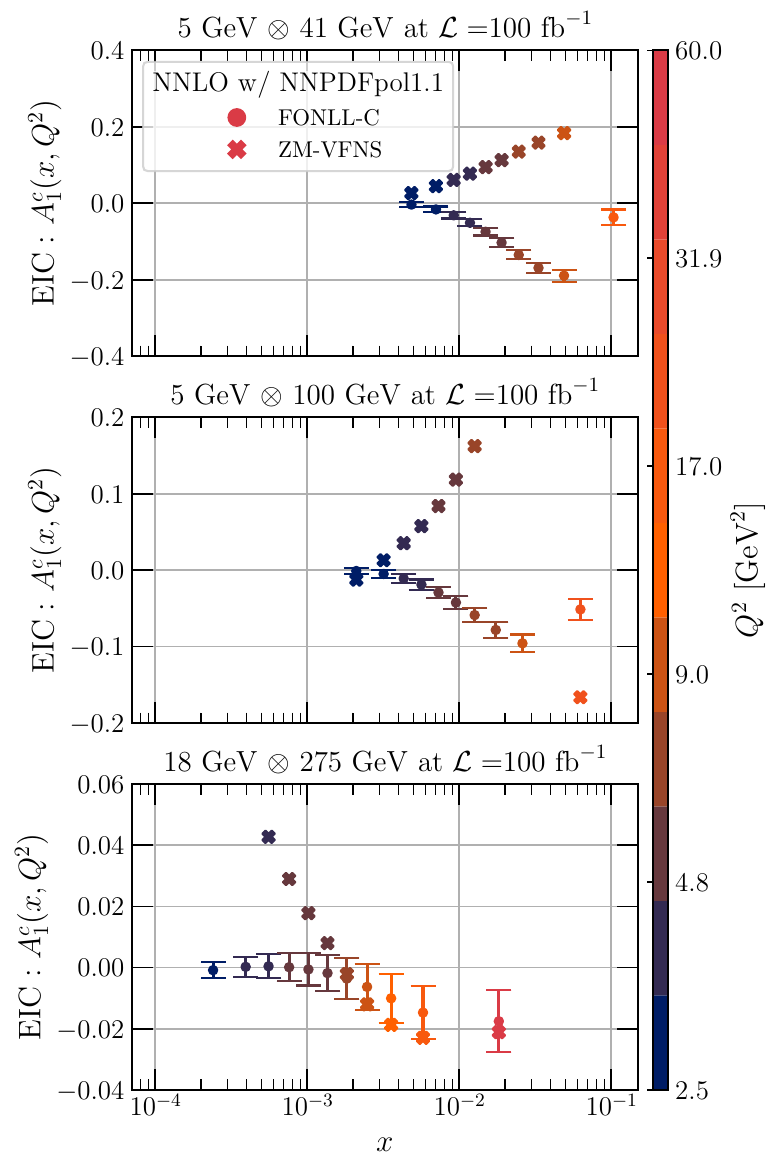}
  \includegraphics[width=0.495\textwidth]{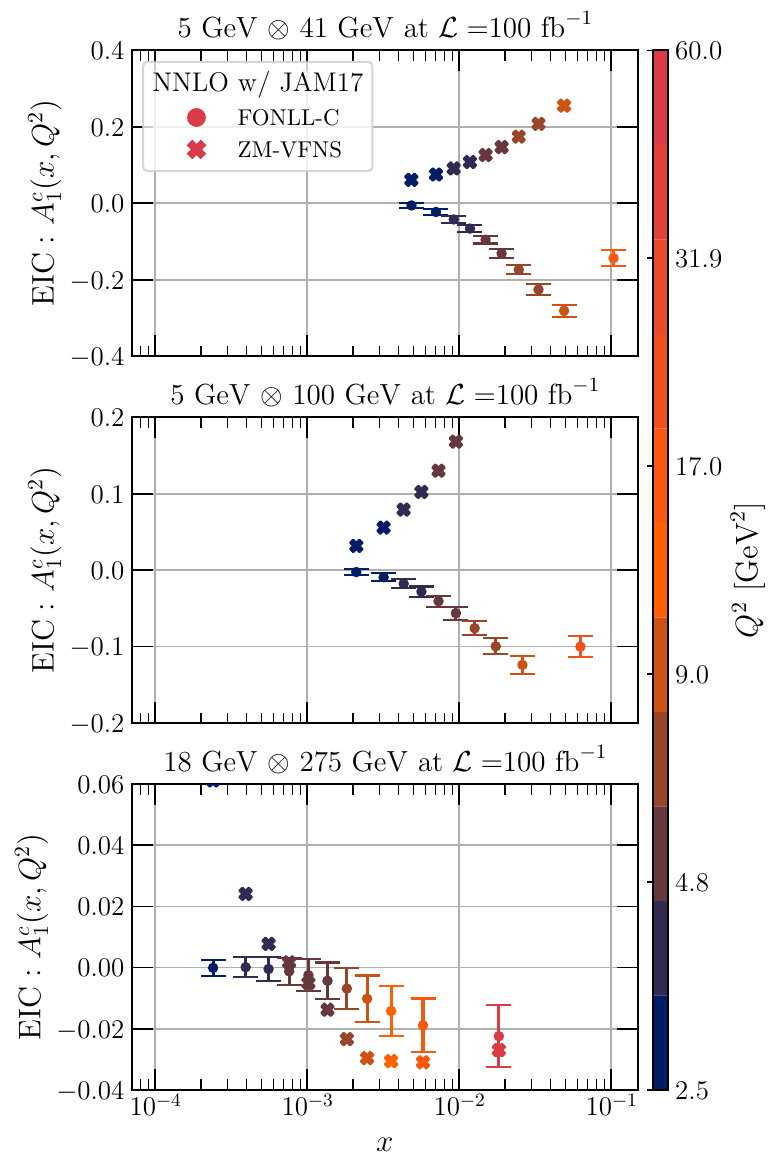}
  \caption{Same as Fig.~\ref{fig:a1c_nlo_EIC} at NNLO.}
  \label{fig:a1c_nnlo_EIC}
\end{figure}

\begin{figure}[!t]
  \centering
  \includegraphics[width=0.495\textwidth]{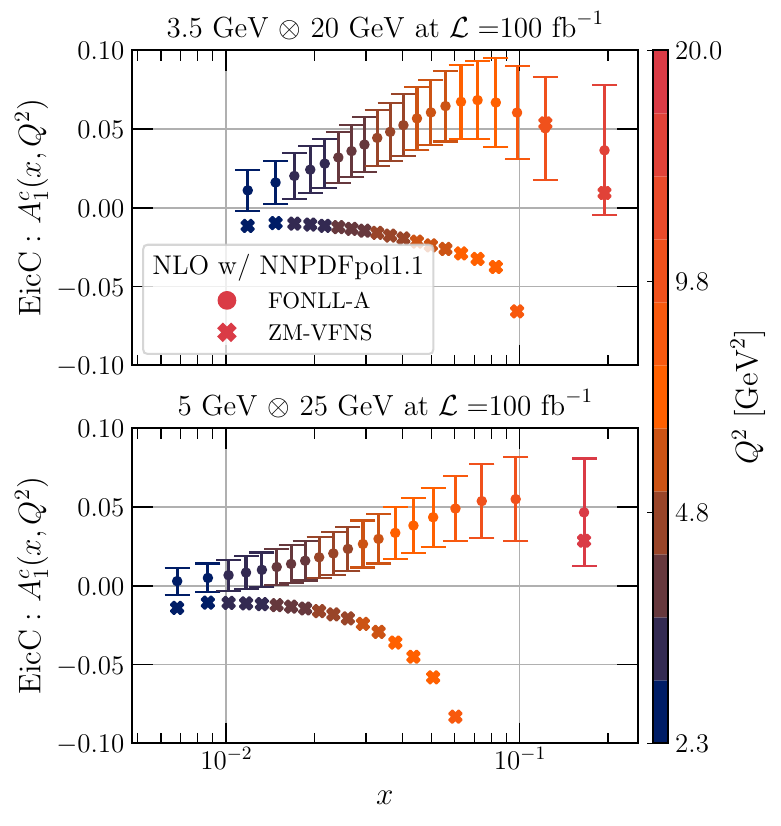}
  \includegraphics[width=0.495\textwidth]{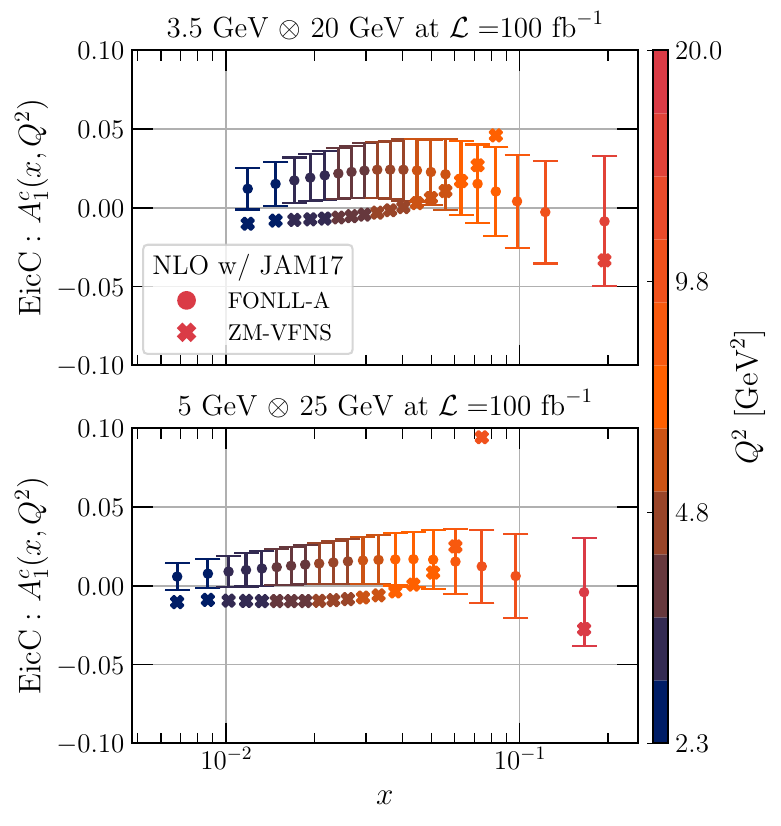}
  \caption{Same as Fig.~\ref{fig:a1c_nlo_EIC} for the EicC.}
  \label{fig:a1c_nlo_EicC}
\end{figure}

\begin{figure}[!t]
  \centering
  \includegraphics[width=0.495\textwidth]{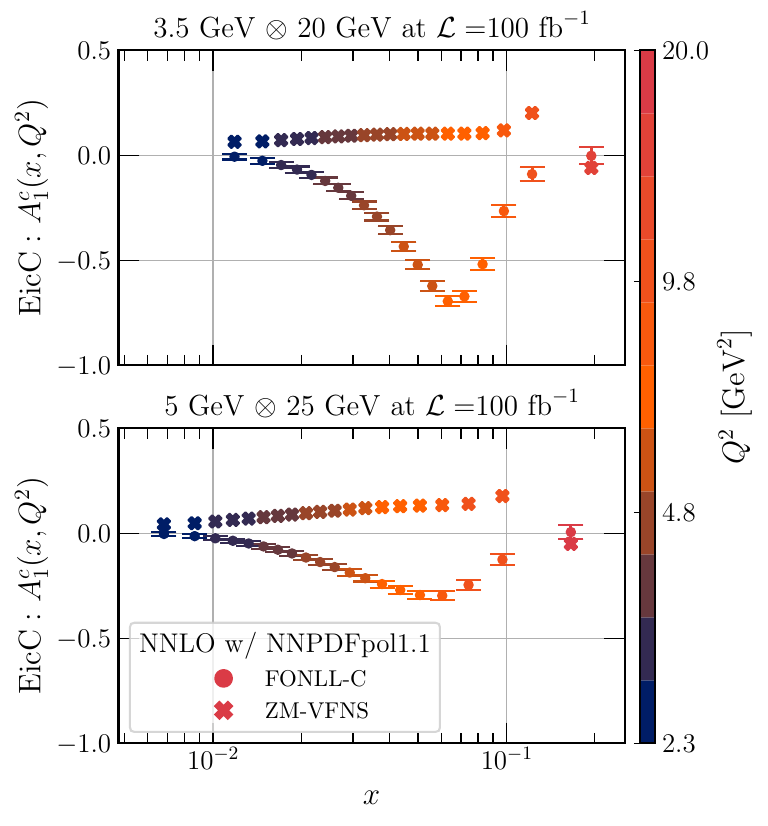}
  \includegraphics[width=0.495\textwidth]{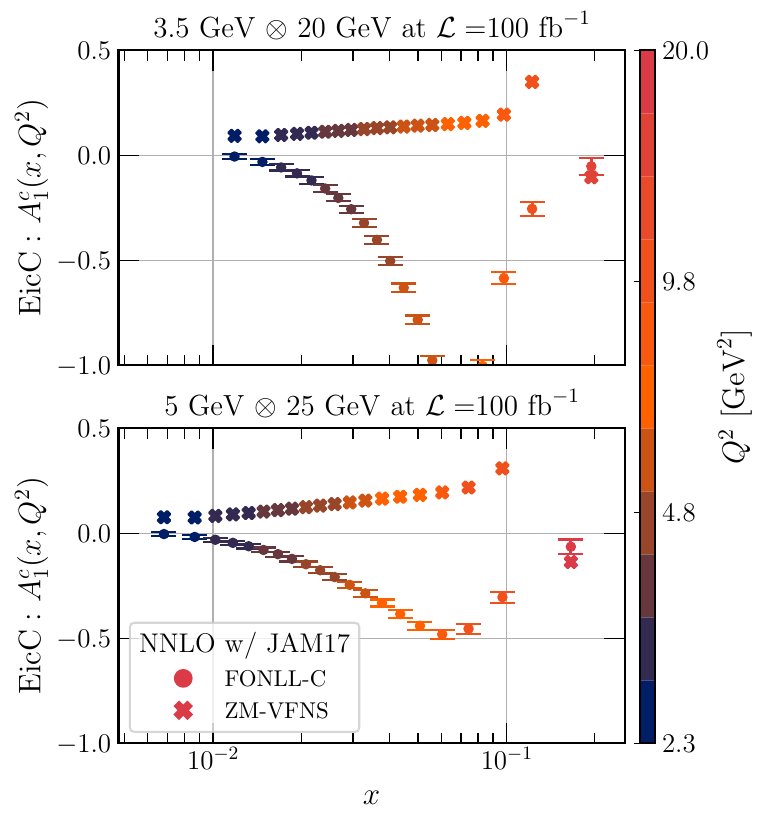}
  \caption{Same as Fig.~\ref{fig:a1c_nlo_EicC} at NNLO.}
  \label{fig:a1c_nnlo_EicC}
\end{figure}

Figures.~\ref{fig:a1c_nlo_EIC} and~\ref{fig:a1c_nnlo_EIC} show the charm
longitudinal asymmetry $A^c_1$, Eq.~\eqref{eq:A1c}, computed at NLO and NNLO
accuracy, respectively. Predictions are obtained either with the ZM-VFN or the
appropriate FONLL schemes (FONLL-A at NLO and FONLL-C at NNLO). They correspond
to the EIC pseudodata discussed in Sect.~\ref{subsec:pseudodata_sets}.
The NNPDFpol1.1 and JAM17 PDF sets are used. Error bars, indicated on top of
the FONLL-C result, correspond to the projected experimental uncertainties.
Figures~\ref{fig:a1c_nlo_EicC} and~\ref{fig:a1c_nnlo_EicC} are as
Figs.~\ref{fig:a1c_nlo_EIC} and~\ref{fig:a1c_nnlo_EIC} for the EicC pseudodata.
In all of these figures, each point corresponds to a different bin 
in $x$ and $Q^2$; due to the DIS kinematics, increasing values of $x$ correlate
with increasing values of $Q^2$.

As in the case of the inclusive double spin asymmetry, we remark that
predictions obtained with either the ZM-VFN of the FONLL schemes, given a
perturbative order, may differ significantly. Differences, as expected,
are generally larger when $Q^2$ is smaller. As in the case of the inclusive
double-spin asymmetry, these are fairly independent from the input PDF set,
and can be larger than the projected experimental uncertainty.
We therefore conclude, also in this case, that the inclusion of charm mass
corrections is essential to correctly interpret future collider data.

We finally note that, for both the EIC and EicC, there are marked differences
in predictions obtained at NLO and NNLO. These can be traced back to the large
perturbative corrections that affect the polarised charm structure function
$g_1^c$ at low $Q^2$, and, albeit to a lesser extent, also its unpolarised
counterpart $F_1^c$. For instance, at $x\sim 0.01$ and $Q^2\sim 5$ GeV$^2$, one
has (using NNPDFpol1.1 as input) that $g_1^c\sim 0.004$ with FONLL-A but
$g_1^c\sim -0.05$ with FONLL-C (Fig.~\ref{fig:g1_charm_pto}):
not only a change of an order of magnitude in size but also a change of sign.
These large perturbative corrections to charm production in polarised
electron-proton collisions are also relevant for the inclusive structure
function, which is reduced from $g_1\sim0.47$ at NLO to $\sim 0.23$ at NNLO
(Fig.~\ref{fig:g1_total_pto}), again considering $x\sim 0.01$ and 
$Q^2\sim 5$ GeV$^2$.

In light of all of these considerations, we generally remark that the
intermediate-to-large-$Q^2$, large-$x$ region, especially for the higher
energy beam configurations, are the most promising to measure a non-vanishing
polarised charm asymmetry, which may be as large 
as a few percent. Such a sizeable asymmetry will provide valuable
information on both the 
mechanisms of heavy quark production in polarised DIS, as well as on the
underlying distribution of the proton spin among its partonic constituents.

\section{Summary}
\label{sec:summary}

In this work we have presented a comprehensive framework enabling the
calculation of polarised structure functions and asymmetries in deep-inelastic
scattering up to $\mathcal{O}\lp \alpha_s^2\rp$ and accounting for charm quark
mass effects. This framework mirrors state-of-the-art 
theory calculations in polarised DIS and is implemented in the open-source
{\sc\small EKO} and {\sc\small YADISM} software. We have shown that FONLL
structure functions successfully match the massless and massive calculations,
and that they display good perturbative convergence.
By comparing our predictions with projected pseudodata
corresponding to the upcoming US- and China-based electron-ion colliders,
we have found that charm mass effects are significant and must be accounted for
to achieve a robust description of both inclusive and charm-tagged polarised
asymmetries at these future facilities.

Our results constitute the first step towards a new global determination of
polarised PDFs accurate to NNLO within the NNPDF framework.
This will possibly include not only polarised DIS measurements, but also 
$W$ gauge boson production and semi-inclusive DIS measurements, for which NNLO
computations have been completed 
recently~\cite{Boughezal:2021wjw,Abele:2021nyo,Goyal:2023xfi}.
Aside from this goal, 
our results represent an important ingredient for the precision phenomenology
program at the upcoming EIC, making it possible to robustly access precious
information on the spin structure of the proton from the interpretation of its
inclusive and charm-tagged polarised structure function measurements.

\subsection*{Acknowledgments}

We thank Daniel de Florian, Rodolfo Sassot, Marco Stratmann, and Werner
Vogelsang for sending us the Monte Carlo variant of DSSV14. We thank Yuxiang
Zhao for sharing the projections for EIC and EicC pseudo-data. We thank Barak
Schmookler for useful discussions and for sharing with us the polarised
structure function projections used in the ATHENA proposal. We also thank
Valerio Bertone for discussions regarding the benchmark of the polarised
coefficient functions.

F.~H. is supported by the Academy of Finland
project 358090 and is funded as a part of the Center
of Excellence in Quark Matter of the Academy of Finland, project 346326.
E.R.~N. is supported by the Italian Ministry of University and Research (MUR)
through the “Rita Levi-Montalcini” Program. J.~R. and G.~M. are partially
supported by NWO (Dutch Research Council). J.~R. and T.~R. are supported by an
ASDI (Accelerating Scientific Discoveries) grant from the Netherlands eScience
Center. R.~S. is supported by the U.K. Science and Technology Facility Council
(STFC) grant ST/T000600/1.

\appendix
\section{Polarised DGLAP evolution}
\label{app:dglap}

Helicity-dependent PDFs obey DGLAP evolution equations, which, in $x$ space,
read as
\be
\frac{d}{d \ln Q^2} \Delta f_i(x, Q^2) = \sum_k \int_x^1 \frac{dz}{z} \Delta P_{ik}\left(\frac x z, \alpha_s(Q^2)\right)  \Delta f_k (x, Q^2) \ ,
\label{eq:dglap_pol}
\ee
with $\Delta P_{ik}(x, \alpha_s)$ being the polarised splitting functions and
$i,k$ partonic flavour indexes running over all active quark flavours and the
gluon. DGLAP equations can, for convenience, be expressed in Mellin space
\be
\frac{d}{d \ln Q^2} \Delta f_i(N, Q^2) = - \sum_k\Delta \gamma_{ik}(N, \alpha_s(Q^2))  \Delta f_k (N, Q^2) \ ,
\label{eq:dglap_pol_mellin}
\ee
where convolutions are replaced by products. The quantities
$\Delta \gamma_{ik}$, called polarised anomalous dimensions, are defined by
\be
\Delta \gamma_{ik}(N, \alpha_s(Q^2)) = - \int_0^1 dx\,x^{N-1} \Delta P_{ik}(x, \alpha_s(Q^2)) \, ,
\ee
and likewise for the Mellin space PDFs $\Delta q_k (N, Q^2)$.
Solving the coupled system of Eq.~(\ref{eq:dglap_pol})
or~(\ref{eq:dglap_pol_mellin})
is most efficiently done by rotating to a convenient  flavour basis.
Specifically, one defines the polarised total quark singlet PDF as
\be
\Delta \Sigma(x,Q^2) = \sum_{i=1}^{n_f} \Delta q_i^+(x,Q^2) = \sum_{i=1}^{n_f} \lp  \Delta q_i(x,Q^2)
+ \Delta \bar{q}_i(x,Q^2)\rp \, ,
\ee
which evolves coupled with the polarised gluon $\Delta g$,
while all other quark combinations evolve independently in terms of non-singlet
evolution equations.

The polarised splitting functions $\Delta P_{ik}$ can be evaluated in
perturbative QCD,
\be
\Delta P_{ik}(x,\alpha_s(Q^2)) = \sum_{n=0}^m \alpha_s^{n+1}(Q^2) \Delta P_{ik}^{(n)}(x) \, . 
\label{eq:splitting_pol}
\ee
The complete set of $\Delta P_{ik}$ has been computed at NLO
in~\cite{Gluck:1995yr} and then at NNLO in~\cite{Moch:2014sna,Moch:2015usa,Blumlein:2021ryt}.
At leading order, the polarised quark-to-quark splitting function is identical
to its unpolarised counterpart, $\Delta P_{qq}^{(0)} = P_{qq}^{(0)}$.
Symmetry considerations imply that polarised non-singlet splitting functions
coincide with the spin-averaged ones to all orders after they are
swapped as follows:
\be
\Delta P_{{\rm NS},\pm}^{(n)}(x) = P_{{\rm NS},\mp}^{(n)}(x) \qquad \forall\, n \, .
\ee
Furthermore, helicity conservation implies that the first moment of the 
gluon-to-quark splitting function vanishes
\be
\int_0^{1} x \Delta P_{qg}(x,\alpha_s(Q^2)) dx = 0 \ ,
\ee
to all orders in perturbation theory.

As in the case of unpolarised DGLAP evolution, one also has to account 
for the fact that the number of active quark flavours $n_f$ depends on the
scale $Q^2$. In a VFN scheme, ignoring intrinsic heavy quark contributions,
heavy flavour polarised PDFs are entirely generated
at the scale $\mu_h$ from matching conditions relating schemes
with $n_f$ and $n_f+1$ active quarks.
At the scale  $Q^2 = \mu_h^2$,
these matching relations take the general form 
\be
\label{eq:polarised_matching}
\Delta f_i^{[n_f+1]} (x, \mu^2_h) = \int_{x}^{1} \frac{dz}{z} \sum_{j=g,q,\bar{q}} \Delta f_j^{[n_f]} 
\left( \frac{x}{z}, \mu_h^2 \right) \Delta K_{ij} \left( z, \alpha_s^{[n_f+1]}(\mu_h^2), \frac{\mu_h^2}{m_h^2} \right) \, ,
\ee
where the sum over $j$ includes only the $n_f$ light quark flavours.
The polarised operator matrix elements $\Delta K_{ij}$ have
been computed up to $\mathcal{O}\lp \alpha_s^2\rp$ (NNLO)
accuracy~\cite{Bierenbaum:2022biv}.
Equation~(\ref{eq:polarised_matching}) can be generalized to the case of
intrinsic heavy quarks being present at scales $Q^2< \mu^2_h$.

Helicity-dependent QCD calculations performed in dimensional regularisation
have to address the issue of the definition of the $\gamma_5$ Dirac matrix in
$d \neq 4$ space-time dimensions, which 
enters through the helicity projection operators.
Usually, an ad-hoc renormalisation scheme, called Larin scheme, 
is defined for convenience and then quantities are mapped
to the M-scheme by means of finite transformation.
This technical point becomes relevant for higher-order QCD
calculations involving polarised partons.

The calculation of polarised structure functions
discussed in Sect.~\ref{sec:nnlo-pol-sfs} requires
polarised PDFs evolved up to NNLO.
For this reason, polarised DGLAP evolution up to this accuracy
has been implemented in {\sc\small EKO}~\cite{Candido:2022tld}.
The necessary VFN scheme matching conditions, which were
computed only very recently, are implemented in a public piece of software for
the first time in this work.

\begin{figure}[!t]
  \centering
  \includegraphics[width=\linewidth]{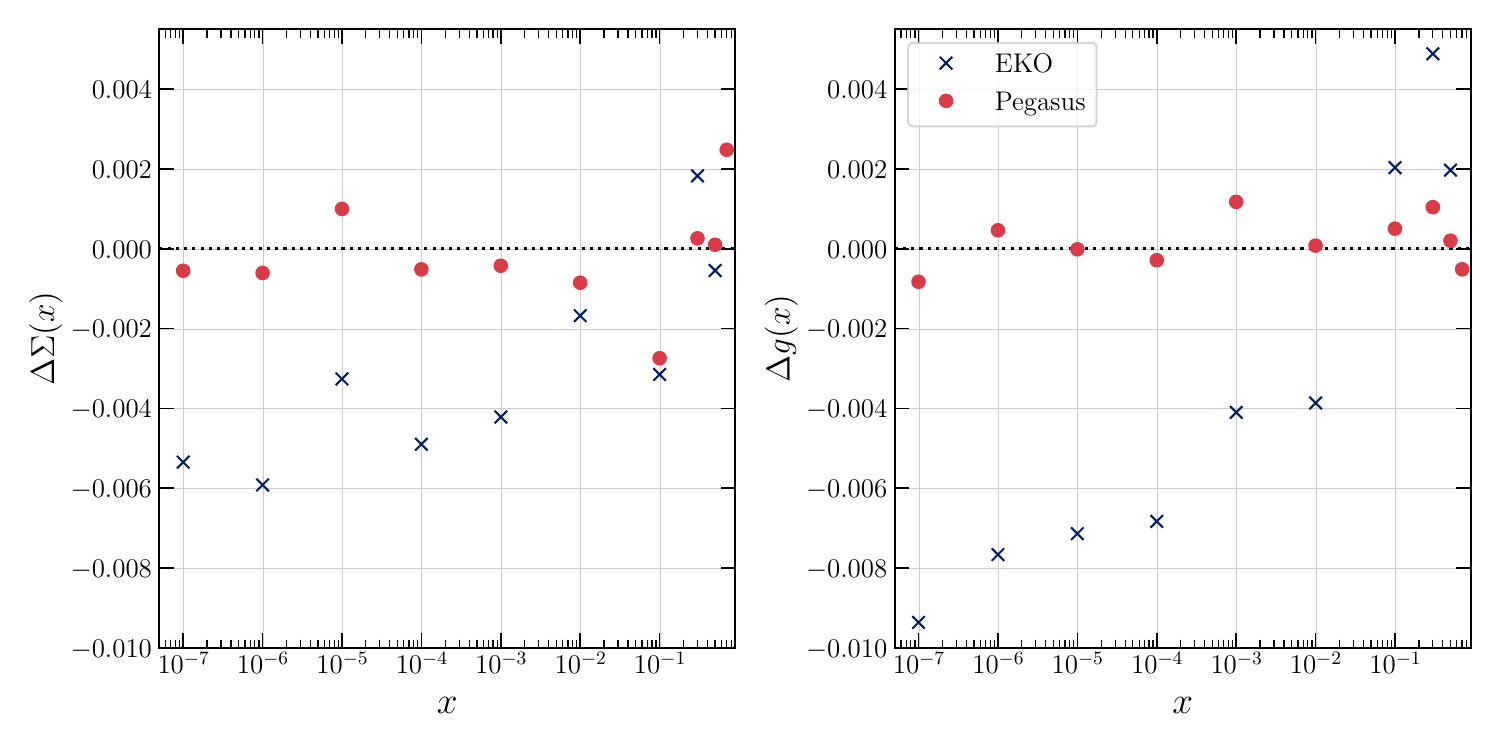}
  \caption{The percentage difference with respect to the Les Houches polarised
    PDF evolution benchmark tables for the
    {\sc\small EKO} and {\sc\small PEGASUS} predictions,
    obtained by evolving
    polarised PDFs from  $Q^2=2$ GeV$^2$ to $Q^2=10^4$ GeV$^2$.
    Evolution is carried out at NLO (the highest accuracy at
    which the polarized
    LH tables are available) in a VFNS with $n_f^{\rm (max)} = 5$ active quark
    flavours.
    We show results for the evolution benchmarks corresponding to the polarised
    quark singlet $\Delta \Sigma(x,Q^2)$ (left)
    and gluon  $\Delta g(x,Q^2)$ PDFs (right). 
  }    
  \label{fig:toylh_benchmark}
\end{figure}

We  have benchmarked our implementation of polarised
evolution in {\sc\small EKO}~ with other public codes, in particular with
{\sc\small PEGASUS}~\cite{Vogt:2004ns}
and {\sc\small APFEL}~\cite{Bertone:2013vaa}, and with the Les Houches (LH)
evolution benchmark tables~\cite{Dittmar:2005ed}, finding good agreement
in all cases considered.
Depending on the case, the benchmark is restricted to the NLO VFN
scheme or to NNLO in the FFN scheme.
In Fig.~\ref{fig:toylh_benchmark} we show the percentage difference
between the results of polarised PDFs evolved from $Q^2=2$ GeV$^2$ to
$Q^2=10^4$ GeV$^2$ with {\sc\small EKO}, {\sc\small PEGASUS}, or the LH tables.
Polarised DGLAP evolution is carried out at NLO (the highest accuracy at
which the LH tables are available) in a VFNS with $n_f^{\rm (max)} = 5$ active
quark flavours. We show results corresponding to the polarised quark singlet
$\Delta \Sigma(x,Q^2)$ and gluon  $\Delta g(x,Q^2)$ PDFs. 
Excellent agreement between the {\sc\small EKO} evolution,
the reference LH tables, and {\sc\small PEGASUS} is found, with differences
at the $\mathcal{O}(10^{-5})$ level.
A similar agreement is achieved for other quark flavour
combinations not shown here.

\begin{figure}[!t]
  \centering
  \includegraphics[width=\linewidth]{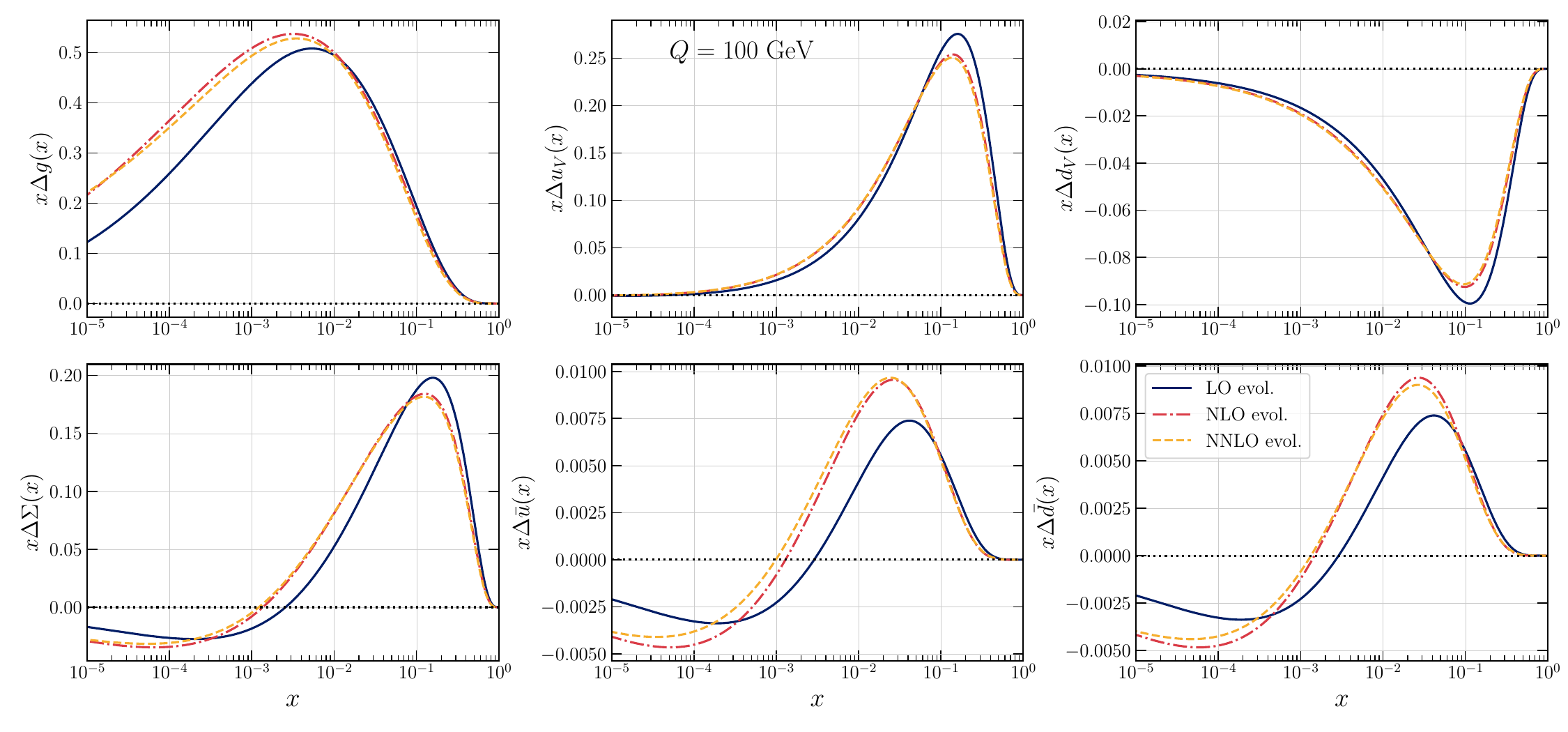}
  \caption{Results of evolving the Les Houches benchmark polarised
    toy PDFs~\cite{Dittmar:2005ed} between $Q^2=2$ GeV$^2$ and $Q^2=10^4$
    GeV$^2$ with {\sc\small EKO} in the VFNS scheme with up to $n_f=5$.
    We compare the results of polarised evolution at three different
    perturbative orders, LO, NLO, and NNLO, for the same fixed input PDF.
  }    
  \label{fig:toy_pdf_evolution}
\end{figure}

We finally assess the perturbative convergence of DGLAP evolution 
for a fixed boundary condition up to this accuracy.
In Fig.~\ref{fig:toy_pdf_evolution} we display the
results of evolving the LH polarised
toy PDFs~\cite{Dittmar:2005ed} between $Q^2=2$ GeV$^2$ 
and $Q^2=10^4$ GeV$^2$ with
{\sc\small EKO} in the VFNS scheme with up to $n_f^{\rm (max)}=5$.
We compare polarised evolution at three different
perturbative orders, LO, NLO, and NNLO, in all cases with the same input PDF.
We note that the LH toy PDF set assumes $\Delta \bar{u} = \Delta \bar{d}$,
while polarised evolution induces a breaking of this relation at NLO and beyond.

From this comparison, one observes how
the perturbative series converges. Differences between
NNLO and NLO are smaller than those between NLO and LO for all values of $x$
and flavour combinations.
NLO corrections are large, for instance up to a factor 2 for the small-$x$
gluon and sea quarks
and up to a 50\% for the sea quarks in the intermediate $x$ region.
NNLO corrections are much smaller, at the few percent level at most.
Although small, these corrections are of the same size or larger as the
projected experimental uncertainties of the forthcoming EIC and EicC
measurements.
Therefore, NNLO corrections to polarised PDF evolution
must be included in theoretical predictions entering the interpretation of the
data at these future facilities.

\section{Target mass corrections in polarised DIS}
\label{sec:tmc-impacts}

In this Appendix, we discuss the implementation of target mass corrections
(TMCs) to the computation of the polarised structure functions $g_1$
in {\sc\small YADISM}.
We also study their numerical impact in comparison to charm quark corrections 
accounted for in the FONLL scheme.

The leading-twist definition of the polarised structure function $g_1$,
Eq.~\eqref{eq:leading-twist-g1}, is valid in the Bjorken-scaling
limit where $Q^2 \to \infty$ and $x$ is fixed. At low $Q^2$ values,
power-suppressed (highest-twist) corrections to the spin-dependent structure
functions can have large effects in some kinematic regions. A subset of the
total higher-twist contribution can be evaluated 
in terms of closed-form expressions using the Operator Product Expansion (OPE).
The complete TMC expressions for polarised structure functions arising from 
twist-2 and twist-3 operators have been derived in~\cite{Blumlein:1998nv}.
Equation~(\ref{eq:leading-twist-g1}) is then modified as
\begin{align}
  \begin{split}
    \tilde{g}_1 \left(x ,Q^2\right) &= \frac{1}{\left(1+\gamma^2\right)^{3 / 2}} \frac{x}{\xi} g_1(\xi,Q^2) 
    \\&+\frac{\gamma^2}{\left(1+\gamma^2\right)^2} \int_{\xi}^1 \frac{d v}{v}\left[\frac{x+\xi}{\xi}
      +\frac{\gamma^2-2}{2 \sqrt{1+\gamma^2}} \log \left(\frac{v}{\xi}\right)\right] g_1 (v,Q^2) \, ,
  \end{split}
  \label{eq:tmc-g1}
\end{align}
in terms of the so-called  Nachtmann variable,
\begin{equation}
  \xi = \frac{2 x}{1+\sqrt{1+\gamma^2}} \quad \text{with} \quad \gamma^2 = 4 x^2 \frac{m_N^2}{Q^2}.
  \label{eq:Nachtmann}
\end{equation}
It is clear from Eq.~(\ref{eq:tmc-g1}) that in the asymptotic limit,
$Q^2 \to \infty$, one has that  $\xi=x$ and
$\gamma=0$ so that the leading-twist expression is recovered. 

\begin{figure}[!t]
  \centering
  \includegraphics[width=1\textwidth]{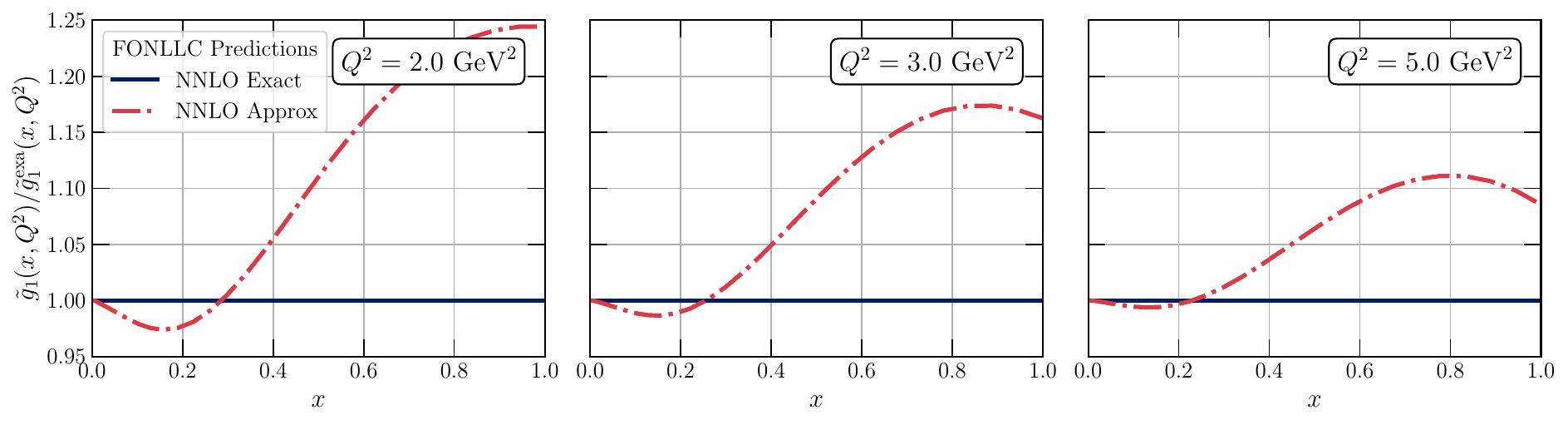}
  \caption{Comparisons of the exact implementation of
    TMCs in $g_1(x,Q^2)$, Eq.~(\ref{eq:tmc-g1}) with the approximated
    implementation of Eq.~(\ref{eq:tmc-g1-approx}) for different values
    of $Q^2$. The polarised structure functions are computed using FONLL-C
    using NNPDFpol1.1 as input PDF set.}
  \label{fig:tmc_approximation}
\end{figure}

The target mass corrected structure function in Eq.~\ref{eq:tmc-g1} involves
integrals, $k_1 = \int dv/v g_1(v)$ and $k_2 = \int dv/v \log (v/ \xi) g_1(v)$,
which can be numerically difficult to evaluate. As done in the
case of unpolarised TMCs~\cite{Schienbein:2007gr}, upper bounds for the size
of these integrals can be computed. Given that the non-leading terms $k_1$
and $k_2$ constitute a small correction to the leading term and that $g_1(v)$
decreases as a function of $v$, we can evaluate the terms at the lower
integration limit. That is, the two integrals have as 
upper bounds $k_1 < g_1 \int dv /v$ and $k_2 < g_1 \int dv/v \log (v/ \xi)$.
By analytically evaluating these two integrals, one arrives at the following
approximation:
\begin{align}
  \begin{split}
    \tilde{g}_1^{\rm approx} \left(x ,Q^2\right) = g_1(\xi,Q^2) \left[ \frac{1}{\left(1+\gamma^2\right)^{3 / 2}} \frac{x}{\xi} + \frac{\gamma^2}
      {\left(1+\gamma^2\right)^2} \left( \frac{x+\xi}{\xi} (1 - \xi) +\frac{\gamma^2-2}{4 \sqrt{1+\gamma^2}}
      \log^2 \left( \frac{1}{\xi} \right) \right) \right],
  \end{split}
  \label{eq:tmc-g1-approx}
\end{align}
with the key benefit that the dependence on the leading-twist structure
function is now factorised.

Figure~\ref{fig:tmc_approximation} compares the TMC-corrected structure
function $\widetilde{g}_1(x,Q^2)$ using the FONLL-C scheme evaluated with the
exact, Eq.~(\ref{eq:tmc-g1}), and  with the approximated,
Eq.~(\ref{eq:tmc-g1-approx}), implementations. These comparisons show that the
difference between the exact and 
approximated expressions is larger at small $Q^2$, reaching up to $25\%$ for 
$Q^2=2~\rm{GeV}^2$, but it decreases as $Q^2$ increases.
Therefore, the target-mass corrected expression given by
Eq.~\ref{eq:tmc-g1-approx} is a 
good approximation at high momentum scale.

\begin{figure}[!t]
  \centering
  \includegraphics[width=1\textwidth]{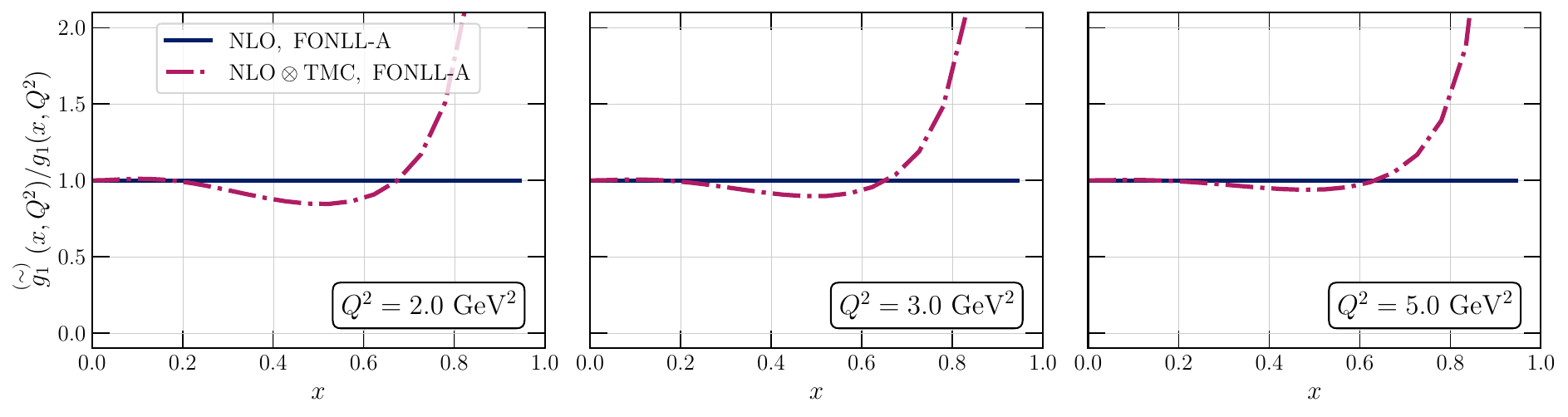}
  \includegraphics[width=1\textwidth]{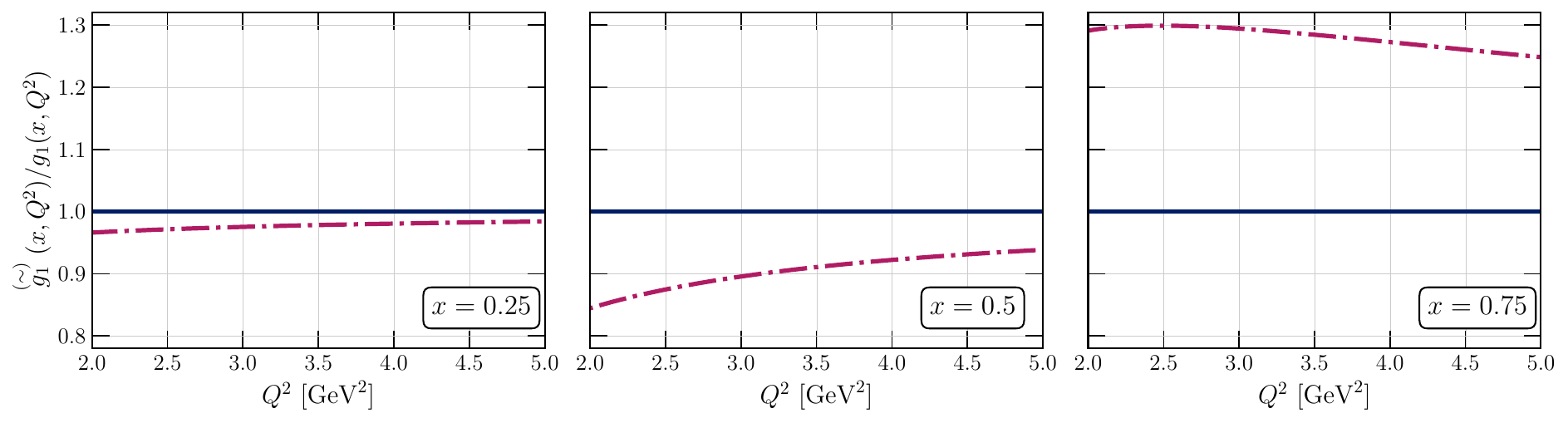}
  \caption{The impact of TMCs on the  $g_1(x,Q^2)$ structure function evaluated
    with FONLL-A as a function of $x$ for fixed $Q^2$ (upper panels)
    and as a function of $Q^2$ for fixed $x$ (lower panels), normalised
    to the calculation without TMCs, using the NNPDFpol1.1 polarised PDF set as
    input.}
  \label{fig:tmc_nlo_nnpdfpol}
\end{figure}

\begin{figure}[!t]
  \centering
  \includegraphics[width=1\textwidth]{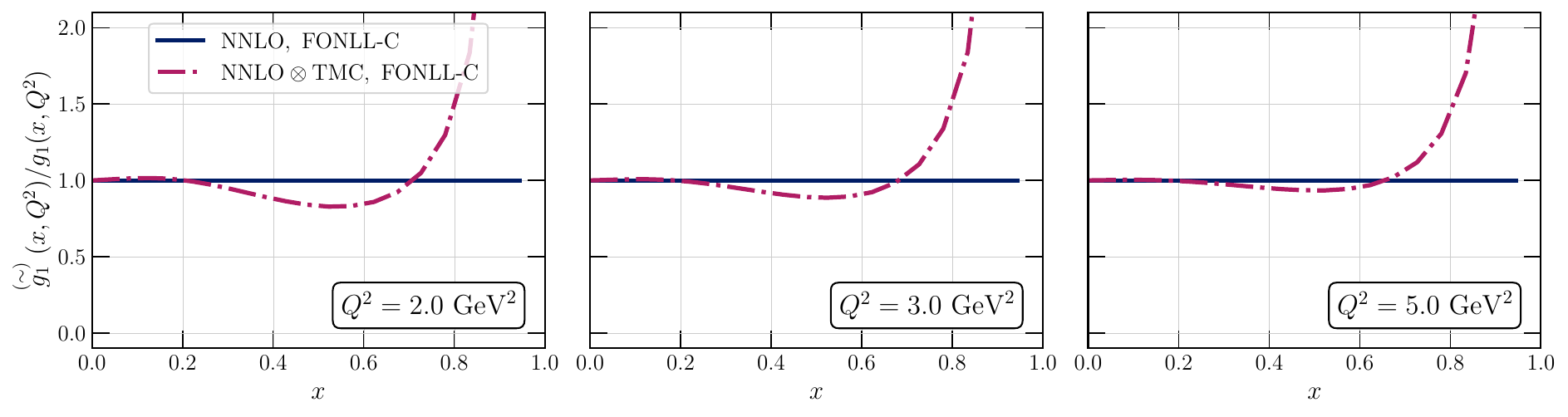}
  \includegraphics[width=1\textwidth]{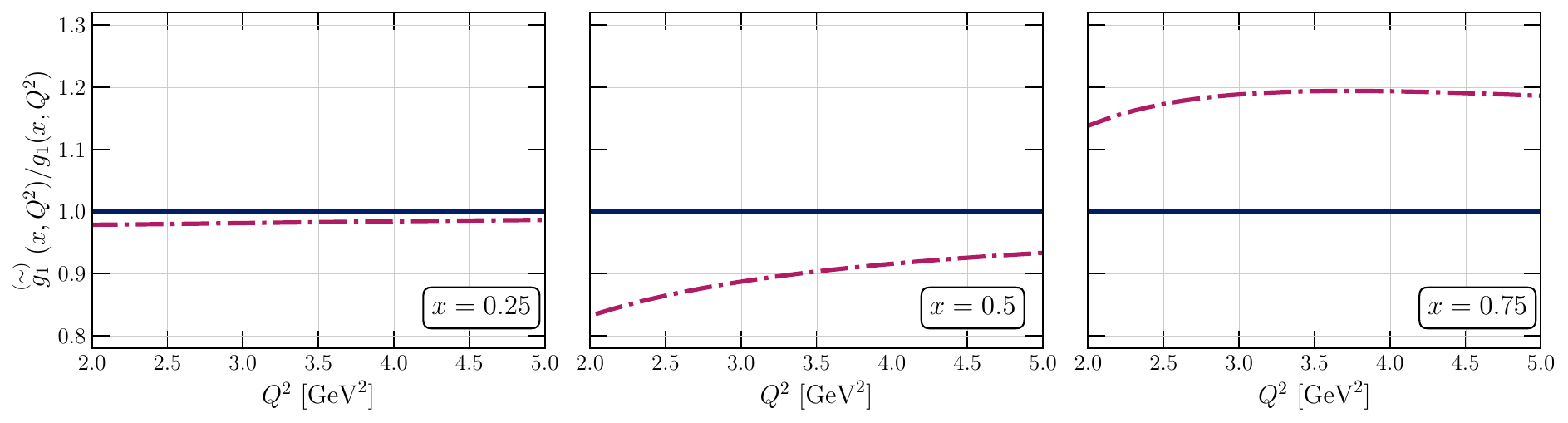}
  \caption{Same as Fig.~\ref{fig:tmc_nlo_nnpdfpol} in the case
    where structure functions are evaluated with FONLL-C.} 
  \label{fig:tmc_nnlo_nnpdfpol}
\end{figure}

The impact of TMCs on the  $g_1(x,Q^2)$ polarised structure function is
displayed in Fig.~\ref{fig:tmc_nlo_nnpdfpol} (with FONLL-A) and in 
Fig.~\ref{fig:tmc_nnlo_nnpdfpol} (with FONLL-C), where the target-mass
corrected structure function is normalised to its leading-twist counterpart.
The NNPDFpol1.1 polarised PDF set is used as input.
As can be seen, the ratio is close to unity at small $x$ and large $Q^2$, while
it quickly increases at large $x$ and large $Q^2$. For instance, at $x=0.5$,
the impact of TMCs on the structure function $g_1$ can grow from about $5 \%$
at $Q^2=5$~GeV$^2$ to about $15\%$ at $Q^2=2$~GeV$^2$.  As expected, these
effects decrease as $x$ 
decreases ($2$-$3\%$ at $x=0.25$) but they increase dramatically as $x$ 
increases ($30\%$ at $x$=0.75). In general, TMCs depend only moderately on the 
perturbative order.

The results of Figs.~\ref{fig:tmc_nlo_nnpdfpol} and~\ref{fig:tmc_nnlo_nnpdfpol}
highlight how TMCs are in general also required to achieve an accurate
description of polarised DIS structure functions, especially for measurements
sensitive to the large-$x$, small-$Q^2$ kinematic region. This region is not
primarily probed by the EIC and the EicC. Furthermore, owing to the fact that
TMCs for the unpolarised structure function $F_1$ and the polarised $g_1$
structure function have similar structure, TMCs mostly cancel in the asymmetry
$A_1$~\cite{Accardi:2008pc}. 
For these reasons, TMCs are not included in the results presented in 
Sects.~\ref{sec:nnlo-pol-sfs}-\ref{sec:fonll-predictions}.

\bibliography{fonll_pol}

\providecommand{\href}[2]{#2}\begingroup\raggedright\begin{thebibliography}{10}

\bibitem{H1:2015ubc}
{\bf H1, ZEUS} Collaboration, H.~Abramowicz et~al., {\it {Combination of
  measurements of inclusive deep inelastic ${e^{\pm }p}$ scattering cross
  sections and QCD analysis of HERA data}},  {\em Eur. Phys. J. C} {\bf 75}
  (2015), no.~12 580, [\href{http://arxiv.org/abs/1506.06042}{{\tt
  arXiv:1506.06042}}].

\bibitem{H1:2018flt}
{\bf H1, ZEUS} Collaboration, H.~Abramowicz et~al., {\it {Combination and QCD
  analysis of charm and beauty production cross-section measurements in deep
  inelastic $ep$ scattering at HERA}},  {\em Eur. Phys. J. C} {\bf 78} (2018),
  no.~6 473, [\href{http://arxiv.org/abs/1804.01019}{{\tt arXiv:1804.01019}}].

\bibitem{Gao:2017yyd}
J.~Gao, L.~Harland-Lang, and J.~Rojo, {\it {The Structure of the Proton in the
  LHC Precision Era}},  {\em Phys. Rept.} {\bf 742} (2018) 1--121,
  [\href{http://arxiv.org/abs/1709.04922}{{\tt arXiv:1709.04922}}].

\bibitem{Kovarik:2019xvh}
K.~Kova\v{r}\'\i{}k, P.~M. Nadolsky, and D.~E. Soper, {\it {Hadronic structure
  in high-energy collisions}},  {\em Rev. Mod. Phys.} {\bf 92} (2020), no.~4
  045003, [\href{http://arxiv.org/abs/1905.06957}{{\tt arXiv:1905.06957}}].

\bibitem{Ethier:2020way}
J.~J. Ethier and E.~R. Nocera, {\it {Parton Distributions in Nucleons and
  Nuclei}},  {\em Ann. Rev. Nucl. Part. Sci.} {\bf 70} (2020) 43--76,
  [\href{http://arxiv.org/abs/2001.07722}{{\tt arXiv:2001.07722}}].

\bibitem{Alekhin:2017kpj}
S.~Alekhin, J.~Bl\"umlein, S.~Moch, and R.~Placakyte, {\it {Parton distribution
  functions, $\alpha_s$, and heavy-quark masses for LHC Run II}},  {\em Phys.
  Rev. D} {\bf 96} (2017), no.~1 014011,
  [\href{http://arxiv.org/abs/1701.05838}{{\tt arXiv:1701.05838}}].

\bibitem{Hou:2019efy}
T.-J. Hou et~al., {\it {New CTEQ global analysis of quantum chromodynamics with
  high-precision data from the LHC}},  {\em Phys. Rev. D} {\bf 103} (2021),
  no.~1 014013, [\href{http://arxiv.org/abs/1912.10053}{{\tt
  arXiv:1912.10053}}].

\bibitem{Bailey:2020ooq}
S.~Bailey, T.~Cridge, L.~A. Harland-Lang, A.~D. Martin, and R.~S. Thorne, {\it
  {Parton distributions from LHC, HERA, Tevatron and fixed target data: MSHT20
  PDFs}},  {\em Eur. Phys. J. C} {\bf 81} (2021), no.~4 341,
  [\href{http://arxiv.org/abs/2012.04684}{{\tt arXiv:2012.04684}}].

\bibitem{NNPDF:2021njg}
{\bf NNPDF} Collaboration, R.~D. Ball et~al., {\it {The path to proton
  structure at 1\% accuracy}},  {\em Eur. Phys. J. C} {\bf 82} (2022), no.~5
  428, [\href{http://arxiv.org/abs/2109.02653}{{\tt arXiv:2109.02653}}].

\bibitem{Alekhin:2012vu}
S.~Alekhin, J.~Bl\"umlein, K.~Daum, K.~Lipka, and S.~Moch, {\it {Precise
  charm-quark mass from deep-inelastic scattering}},  {\em Phys. Lett. B} {\bf
  720} (2013) 172--176, [\href{http://arxiv.org/abs/1212.2355}{{\tt
  arXiv:1212.2355}}].

\bibitem{Thorne:1997uu}
R.~S. Thorne and R.~G. Roberts, {\it {A Practical procedure for evolving heavy
  flavor structure functions}},  {\em Phys. Lett. B} {\bf 421} (1998) 303--311,
  [\href{http://arxiv.org/abs/hep-ph/9711223}{{\tt hep-ph/9711223}}].

\bibitem{Forte:2010ta}
S.~Forte, E.~Laenen, P.~Nason, and J.~Rojo, {\it {Heavy quarks in
  deep-inelastic scattering}},  {\em Nucl. Phys. B} {\bf 834} (2010) 116--162,
  [\href{http://arxiv.org/abs/1001.2312}{{\tt arXiv:1001.2312}}].

\bibitem{Gao:2013wwa}
J.~Gao, M.~Guzzi, and P.~M. Nadolsky, {\it {Charm quark mass dependence in a
  global QCD analysis}},  {\em Eur. Phys. J. C} {\bf 73} (2013), no.~8 2541,
  [\href{http://arxiv.org/abs/1304.3494}{{\tt arXiv:1304.3494}}].

\bibitem{Nocera:2014gqa}
{\bf NNPDF} Collaboration, E.~R. Nocera, R.~D. Ball, S.~Forte, G.~Ridolfi, and
  J.~Rojo, {\it {A first unbiased global determination of polarized PDFs and
  their uncertainties}},  {\em Nucl. Phys. B} {\bf 887} (2014) 276--308,
  [\href{http://arxiv.org/abs/1406.5539}{{\tt arXiv:1406.5539}}].

\bibitem{deFlorian:2014yva}
D.~de~Florian, R.~Sassot, M.~Stratmann, and W.~Vogelsang, {\it {Evidence for
  polarization of gluons in the proton}},  {\em Phys. Rev. Lett.} {\bf 113}
  (2014), no.~1 012001, [\href{http://arxiv.org/abs/1404.4293}{{\tt
  arXiv:1404.4293}}].

\bibitem{DeFlorian:2019xxt}
D.~De~Florian, G.~A. Lucero, R.~Sassot, M.~Stratmann, and W.~Vogelsang, {\it
  {Monte Carlo sampling variant of the DSSV14 set of helicity parton
  densities}},  {\em Phys. Rev. D} {\bf 100} (2019), no.~11 114027,
  [\href{http://arxiv.org/abs/1902.10548}{{\tt arXiv:1902.10548}}].

\bibitem{Ethier:2017zbq}
J.~J. Ethier, N.~Sato, and W.~Melnitchouk, {\it {First simultaneous extraction
  of spin-dependent parton distributions and fragmentation functions from a
  global QCD analysis}},  {\em Phys. Rev. Lett.} {\bf 119} (2017), no.~13
  132001, [\href{http://arxiv.org/abs/1705.05889}{{\tt arXiv:1705.05889}}].

\bibitem{AbdulKhalek:2021gbh}
R.~Abdul~Khalek et~al., {\it {Science Requirements and Detector Concepts for
  the Electron-Ion Collider}: {EIC Yellow Report}},  {\em Nucl. Phys. A} {\bf
  1026} (2022) 122447, [\href{http://arxiv.org/abs/2103.05419}{{\tt
  arXiv:2103.05419}}].

\bibitem{Anderle:2021wcy}
D.~P. Anderle et~al., {\it {Electron-ion collider in China}},  {\em Front.
  Phys. (Beijing)} {\bf 16} (2021), no.~6 64701,
  [\href{http://arxiv.org/abs/2102.09222}{{\tt arXiv:2102.09222}}].

\bibitem{Ball:2015tna}
R.~D. Ball, V.~Bertone, M.~Bonvini, S.~Forte, P.~Groth~Merrild, J.~Rojo, and
  L.~Rottoli, {\it {Intrinsic charm in a matched general-mass scheme}},  {\em
  Phys. Lett. B} {\bf 754} (2016) 49--58,
  [\href{http://arxiv.org/abs/1510.00009}{{\tt arXiv:1510.00009}}].

\bibitem{Candido:2022tld}
A.~Candido, F.~Hekhorn, and G.~Magni, {\it {EKO: evolution kernel operators}},
  {\em Eur. Phys. J. C} {\bf 82} (2022), no.~10 976,
  [\href{http://arxiv.org/abs/2202.02338}{{\tt arXiv:2202.02338}}].

\bibitem{Candido:2024rkr}
A.~Candido, F.~Hekhorn, G.~Magni, T.~R. Rabemananjara, and R.~Stegeman, {\it
  {Yadism: Yet Another Deep-Inelastic Scattering Module}},
  \href{http://arxiv.org/abs/2401.15187}{{\tt arXiv:2401.15187}}.

\bibitem{Vogt:2004ns}
A.~Vogt, {\it {Efficient evolution of unpolarized and polarized parton
  distributions with QCD-PEGASUS}},  {\em Comput. Phys. Commun.} {\bf 170}
  (2005) 65--92, [\href{http://arxiv.org/abs/hep-ph/0408244}{{\tt
  hep-ph/0408244}}].

\bibitem{Zijlstra:1993sh}
E.~B. Zijlstra and W.~L. van Neerven, {\it {Order-$\alpha_s^2$ corrections to
  the polarized structure function $g_1 (x,Q^2)$}},  {\em Nucl. Phys. B} {\bf
  417} (1994) 61--100. [Erratum: Nucl.Phys.B 426, 245 (1994), Erratum:
  Nucl.Phys.B 773, 105--106 (2007), Erratum: Nucl.Phys.B 501, 599--599 (1997)].

\bibitem{Blumlein:2022gpp}
J.~Bl\"umlein, P.~Marquard, C.~Schneider, and K.~Sch\"onwald, {\it {The
  massless three-loop Wilson coefficients for the deep-inelastic structure
  functions F$_{2}$, F$_{L}$, xF$_{3}$ and g$_{1}$}},  {\em JHEP} {\bf 11}
  (2022) 156, [\href{http://arxiv.org/abs/2208.14325}{{\tt arXiv:2208.14325}}].

\bibitem{deFlorian:1994wp}
D.~de~Florian and R.~Sassot, {\it {O (alpha-s) spin dependent weak structure
  functions}},  {\em Phys. Rev. D} {\bf 51} (1995) 6052--6058,
  [\href{http://arxiv.org/abs/hep-ph/9412255}{{\tt hep-ph/9412255}}].

\bibitem{Anselmino:1996cd}
M.~Anselmino, P.~Gambino, and J.~Kalinowski, {\it {New proton polarized
  structure functions in charged current processes at HERA}},  {\em Phys. Rev.
  D} {\bf 55} (1997) 5841--5844,
  [\href{http://arxiv.org/abs/hep-ph/9607427}{{\tt hep-ph/9607427}}].

\bibitem{Borsa:2022irn}
I.~Borsa, D.~de~Florian, and I.~Pedron, {\it {The full set of polarized deep
  inelastic scattering structure functions at NNLO accuracy}},  {\em Eur. Phys.
  J. C} {\bf 82} (2022), no.~12 1167,
  [\href{http://arxiv.org/abs/2210.12014}{{\tt arXiv:2210.12014}}].

\bibitem{Hekhorn:2018ywm}
F.~Hekhorn and M.~Stratmann, {\it {Next-to-Leading Order QCD Corrections to
  Inclusive Heavy-Flavor Production in Polarized Deep-Inelastic Scattering}},
  {\em Phys. Rev. D} {\bf 98} (2018), no.~1 014018,
  [\href{http://arxiv.org/abs/1805.09026}{{\tt arXiv:1805.09026}}].

\bibitem{Behring:2015zaa}
A.~Behring, J.~Bl\"umlein, A.~De~Freitas, A.~von Manteuffel, and C.~Schneider,
  {\it {The 3-Loop Non-Singlet Heavy Flavor Contributions to the Structure
  Function $g_{1}(x,Q^{2})$ at Large Momentum Transfer}},  {\em Nucl. Phys. B}
  {\bf 897} (2015) 612--644, [\href{http://arxiv.org/abs/1504.08217}{{\tt
  arXiv:1504.08217}}].

\bibitem{Ablinger:2019etw}
J.~Ablinger, A.~Behring, J.~Bl\"umlein, A.~De~Freitas, A.~von Manteuffel,
  C.~Schneider, and K.~Sch\"onwald, {\it {The three-loop single mass polarized
  pure singlet operator matrix element}},  {\em Nucl. Phys. B} {\bf 953} (2020)
  114945, [\href{http://arxiv.org/abs/1912.02536}{{\tt arXiv:1912.02536}}].

\bibitem{Behring:2021asx}
A.~Behring, J.~Bl\"umlein, A.~De~Freitas, A.~von Manteuffel, K.~Sch\"onwald,
  and C.~Schneider, {\it {The polarized transition matrix element $A_{gq}(N)$
  of the variable flavor number scheme at $O(\alpha^3_s)$}},  {\em Nucl. Phys.
  B} {\bf 964} (2021) 115331, [\href{http://arxiv.org/abs/2101.05733}{{\tt
  arXiv:2101.05733}}].

\bibitem{Blumlein:2021xlc}
J.~Bl\"umlein, A.~De~Freitas, M.~Saragnese, C.~Schneider, and K.~Sch\"onwald,
  {\it {Logarithmic contributions to the polarized O(\ensuremath{\alpha}s3)
  asymptotic massive Wilson coefficients and operator matrix elements in deeply
  inelastic scattering}},  {\em Phys. Rev. D} {\bf 104} (2021), no.~3 034030,
  [\href{http://arxiv.org/abs/2105.09572}{{\tt arXiv:2105.09572}}].

\bibitem{Bierenbaum:2022biv}
I.~Bierenbaum, J.~Bl\"umlein, A.~De~Freitas, A.~Goedicke, S.~Klein, and
  K.~Sch\"onwald, {\it {$O(\alpha_s^2)$ polarized heavy flavor corrections to
  deep-inelastic scattering at $Q^2\gg m^2$}},  {\em Nucl. Phys. B} {\bf 988}
  (2023) 116114, [\href{http://arxiv.org/abs/2211.15337}{{\tt
  arXiv:2211.15337}}].

\bibitem{Ablinger:2023ahe}
J.~Ablinger, A.~Behring, J.~Bl\"umlein, A.~De~Freitas, A.~von Manteuffel,
  C.~Schneider, and K.~Sch\"onwald, {\it {The first-order factorizable
  contributions to the three-loop massive operator matrix elements
  $A_{Qg}^{(3)}$ and $\Delta A_{Qg}^{(3)}$}},
  \href{http://arxiv.org/abs/2311.00644}{{\tt arXiv:2311.00644}}.

\bibitem{Ball:2022qks}
{\bf NNPDF} Collaboration, R.~D. Ball, A.~Candido, J.~Cruz-Martinez, S.~Forte,
  T.~Giani, F.~Hekhorn, K.~Kudashkin, G.~Magni, and J.~Rojo, {\it {Evidence for
  intrinsic charm quarks in the proton}},  {\em Nature} {\bf 608} (2022),
  no.~7923 483--487, [\href{http://arxiv.org/abs/2208.08372}{{\tt
  arXiv:2208.08372}}].

\bibitem{Bertone:2013vaa}
{\bf APFEL} Collaboration, V.~Bertone, S.~Carrazza, and J.~Rojo, {\it {APFEL: A
  PDF Evolution Library with QED corrections}},  {\em Comput. Phys. Commun.}
  {\bf 185} (2014) 1647--1668, [\href{http://arxiv.org/abs/1310.1394}{{\tt
  arXiv:1310.1394}}].

\bibitem{Bertone:2017gds}
V.~Bertone, {\it {APFEL++: A new PDF evolution library in C++}},  {\em PoS}
  {\bf DIS2017} (2018) 201, [\href{http://arxiv.org/abs/1708.00911}{{\tt
  arXiv:1708.00911}}].

\bibitem{Cacciari:1998it}
M.~Cacciari, M.~Greco, and P.~Nason, {\it {The $p_T$ spectrum in heavy-flavour
  hadroproduction.}},  {\em JHEP} {\bf 05} (1998) 007,
  [\href{http://arxiv.org/abs/hep-ph/9803400}{{\tt hep-ph/9803400}}].

\bibitem{Chetyrkin:1997sg}
K.~G. Chetyrkin, B.~A. Kniehl, and M.~Steinhauser, {\it {Strong coupling
  constant with flavor thresholds at four loops in the MS scheme}},  {\em Phys.
  Rev. Lett.} {\bf 79} (1997) 2184--2187,
  [\href{http://arxiv.org/abs/hep-ph/9706430}{{\tt hep-ph/9706430}}].

\bibitem{ATHENA:2022hxb}
{\bf ATHENA} Collaboration, J.~Adam et~al., {\it {ATHENA detector proposal
  \textemdash{} a totally hermetic electron nucleus apparatus proposed for IP6
  at the Electron-Ion Collider}},  {\em JINST} {\bf 17} (2022), no.~10 P10019,
  [\href{http://arxiv.org/abs/2210.09048}{{\tt arXiv:2210.09048}}].

\bibitem{Anderle:2021hpa}
D.~P. Anderle, X.~Dong, F.~Hekhorn, M.~Kelsey, S.~Radhakrishnan,
  E.~Sichtermann, L.~Xia, H.~Xing, F.~Yuan, and Y.~Zhao, {\it {Probing gluon
  helicity with heavy flavor at the Electron-Ion Collider}},  {\em Phys. Rev.
  D} {\bf 104} (2021), no.~11 114039,
  [\href{http://arxiv.org/abs/2110.04489}{{\tt arXiv:2110.04489}}].

\bibitem{Anderle:2023uvi}
D.~P. Anderle, A.~Guo, F.~Hekhorn, Y.~Liang, Y.~Ma, L.~Xia, H.~Xing, and
  Y.~Zhao, {\it {Probing gluon distributions with $D^0$ production at the
  EicC}},  \href{http://arxiv.org/abs/2307.16135}{{\tt arXiv:2307.16135}}.

\bibitem{Ball:2013tyh}
{\bf NNPDF} Collaboration, R.~D. Ball, S.~Forte, A.~Guffanti, E.~R. Nocera,
  G.~Ridolfi, and J.~Rojo, {\it {Polarized Parton Distributions at an
  Electron-Ion Collider}},  {\em Phys. Lett. B} {\bf 728} (2014) 524--531,
  [\href{http://arxiv.org/abs/1310.0461}{{\tt arXiv:1310.0461}}].

\bibitem{Boughezal:2021wjw}
R.~Boughezal, H.~T. Li, and F.~Petriello, {\it {$W$-boson production in
  polarized proton-proton collisions at RHIC through next-to-next-to-leading
  order in perturbative QCD}},  {\em Phys. Lett. B} {\bf 817} (2021) 136333,
  [\href{http://arxiv.org/abs/2101.02214}{{\tt arXiv:2101.02214}}].

\bibitem{Abele:2021nyo}
M.~Abele, D.~de~Florian, and W.~Vogelsang, {\it {Approximate NNLO QCD
  corrections to semi-inclusive DIS}},  {\em Phys. Rev. D} {\bf 104} (2021),
  no.~9 094046, [\href{http://arxiv.org/abs/2109.00847}{{\tt
  arXiv:2109.00847}}].

\bibitem{Goyal:2023xfi}
S.~Goyal, S.-O. Moch, V.~Pathak, N.~Rana, and V.~Ravindran, {\it {NNLO QCD
  corrections to semi-inclusive DIS}},
  \href{http://arxiv.org/abs/2312.17711}{{\tt arXiv:2312.17711}}.

\bibitem{Gluck:1995yr}
M.~Gluck, E.~Reya, M.~Stratmann, and W.~Vogelsang, {\it {Next-to-leading order
  radiative parton model analysis of polarized deep inelastic lepton - nucleon
  scattering}},  {\em Phys. Rev. D} {\bf 53} (1996) 4775--4786,
  [\href{http://arxiv.org/abs/hep-ph/9508347}{{\tt hep-ph/9508347}}].

\bibitem{Moch:2014sna}
S.~Moch, J.~A.~M. Vermaseren, and A.~Vogt, {\it {The Three-Loop Splitting
  Functions in QCD: The Helicity-Dependent Case}},  {\em Nucl. Phys. B} {\bf
  889} (2014) 351--400, [\href{http://arxiv.org/abs/1409.5131}{{\tt
  arXiv:1409.5131}}].

\bibitem{Moch:2015usa}
S.~Moch, J.~A.~M. Vermaseren, and A.~Vogt, {\it {On \ensuremath{\gamma}5 in
  higher-order QCD calculations and the NNLO evolution of the polarized valence
  distribution}},  {\em Phys. Lett. B} {\bf 748} (2015) 432--438,
  [\href{http://arxiv.org/abs/1506.04517}{{\tt arXiv:1506.04517}}].

\bibitem{Blumlein:2021ryt}
J.~Bl\"umlein, P.~Marquard, C.~Schneider, and K.~Sch\"onwald, {\it {The
  three-loop polarized singlet anomalous dimensions from off-shell operator
  matrix elements}},  {\em JHEP} {\bf 01} (2022) 193,
  [\href{http://arxiv.org/abs/2111.12401}{{\tt arXiv:2111.12401}}].

\bibitem{Dittmar:2005ed}
M.~Dittmar et~al., {\it {Working Group I: Parton distributions: Summary report
  for the HERA LHC Workshop Proceedings}},
  \href{http://arxiv.org/abs/hep-ph/0511119}{{\tt hep-ph/0511119}}.

\bibitem{Blumlein:1998nv}
J.~Blumlein and A.~Tkabladze, {\it {Target mass corrections for polarized
  structure functions and new sum rules}},  {\em Nucl. Phys. B} {\bf 553}
  (1999) 427--464, [\href{http://arxiv.org/abs/hep-ph/9812478}{{\tt
  hep-ph/9812478}}].

\bibitem{Schienbein:2007gr}
I.~Schienbein et~al., {\it {A Review of Target Mass Corrections}},  {\em J.
  Phys. G} {\bf 35} (2008) 053101, [\href{http://arxiv.org/abs/0709.1775}{{\tt
  arXiv:0709.1775}}].

\bibitem{Accardi:2008pc}
A.~Accardi and W.~Melnitchouk, {\it {Target mass corrections for spin-dependent
  structure functions in collinear factorization}},  {\em Phys. Lett. B} {\bf
  670} (2008) 114--118, [\href{http://arxiv.org/abs/0808.2397}{{\tt
  arXiv:0808.2397}}].

\end{thebibliography}\endgroup

\end{document}